%% file: arXiv_New.tex
\documentclass[10pt,letterpaper,notitlepage]{article}

\usepackage[ruled]{algorithm2e}

\usepackage{hyperref}
\usepackage{url}
\usepackage{multirow}
\usepackage{hhline}
\usepackage{xcolor,colortbl}
\usepackage{algorithmic}
\usepackage{amsmath}
\usepackage{amssymb}
\usepackage{enumitem}
\usepackage{epsfig}

   \hypersetup{allcolors=blue,colorlinks=true,%
               linkcolor=blue,citecolor=blue,filecolor=blue,urlcolor=blue,colorlinks=true}

\newcommand{\eg}{{\it e.g., }}

\newcommand{\etal}{{\it et~al., }}
\newcommand{\ie}{{\it i.e., }}
\newcommand{\etc}{{\it etc.}}
\newcommand{\ci}{{\it (i) }}
\newcommand{\cii}{{\it (ii) }}
\newcommand{\ciii}{{\it (iii) }}
\newcommand{\ca}{{\it (a) }}
\newcommand{\cb}{{\it (b) }}
\newcommand{\cc}{{\it (c) }}
\newcommand{\cd}{{\it (d) }}

\usepackage{setspace}

\usepackage[T1]{fontenc}
\usepackage[utf8]{inputenc}
\usepackage{authblk}

\begin{document}

% Title portion
\title{\bf \Large Learning Execution Contexts from System Call Distributions for Intrusion Detection in Embedded Systems\thanks{This work is supported in part by grants from NSF CNS 13-02563, NSF CNS 12-19064, NSF CNS 14-23334 and Navy N00014-12-1-0046. Man-Ki Yoon was also supported by Qualcomm Innovation Fellowship and Intel PhD Fellowship. Jaesik Choi was supported by Basic Science Research Program through the National Research Foundation of Korea (NRF) grant funded by the Ministry of Science, ICT \& Future Planning (MSIP) (No. NRF-2014R1A1A1002662 and No. NRF-2014M2A8A2074096). Any opinions, findings, and conclusions or recommendations expressed here are those of the authors and do not necessarily reflect the views of sponsors.
}}

\author[1]{Man-Ki Yoon}
\author[2]{Sibin Mohan}
\author[3]{Jaesik Choi}
\author[4]{Mihai Christodorescu}
\author[1]{Lui Sha}
\affil[1]{\normalsize Department of Computer Science, University of Illinois at Urbana-Champaign}
\affil[2]{Information Trust Institute, University of Illinois at Urbana-Champaign}
\affil[3]{School of Electrial and Computer Engineering, Ulsan National Institute of Science and Technology}
\affil[4]{Qualcomm Research Silicon Valley}
\affil[ ]{ }
\affil[ ]{Email: \{mkyoon, sibin, lrs\}@illinois.edu, jaesik@unist.ac.kr, mihai@qti.qualcomm.com}

\renewcommand\Authands{ and }

%\author{Man-Ki Yoon
%\affil{University of Illinois at Urbana-Champaign}
%Sibin Mohan
%\affil{University of Illinois at Urbana-Champaign}
%Jaesik Choi
%\affil{Ulsan National Institute of Science and Technology}
%Mihai Christodorescu
%\affil{Qualcomm Research Silicon Valley}
%Lui Sha
%\affil{University of Illinois at Urbana-Champaign}}
% NOTE! Affiliations placed here should be for the institution where the
%       BULK of the research was done. If the author has gone to a new
%       institution, before publication, the (above) affiliation should NOT be changed.
%       The authors 'current' address may be given in the "Author's addresses:" block (below).
%       So for example, Mr. Abdelzaher, the bulk of the research was done at UIUC, and he is
%       currently affiliated with NASA.

\date{}

\maketitle

\begin{abstract}
\input{abstract}
\end{abstract}

\input{intro}
\input{overview}
\input{Analysis}
\input{architecture}
\input{Evaluation}
\input{related}
\input{concl}

%\appendixhead{Yoon}

% Acknowledgments
%\begin{acks}
%\end{acks}

% Bibliography
%\bibliographystyle{ACM-Reference-Format-Journals}
%\bibliographystyle{acmtrans}
%\bibliography{acmsmall-sample-bibfile}

\setstretch{0.97}
\bibliographystyle{abbrv}
\bibliography{securecore,sibin.security}

\setstretch{1}
\include{appendix}

\end{document}

%% file: abstract.tex
Existing techniques used for intrusion detection do not fully utilize the intrinsic properties of embedded systems. In this paper, we propose a lightweight method for detecting anomalous executions using a \emph{distribution of system call frequencies}. We use a cluster analysis to learn the legitimate execution contexts of embedded applications and then monitor them at run-time to capture abnormal executions. We also present an architectural framework with minor processor modifications to aid in this process. Our prototype shows that the proposed method can effectively detect anomalous executions without relying on sophisticated analyses or affecting the critical execution paths.

%% file: intro.tex
\section{Introduction}
\label{sec::intro}

An increasing number of attacks are targeting embedded systems~\cite{auto:koscher2010,DroneHack:Shepard2012} that compromise the security, and hence safety, of such systems. %The ever-growing complexity of modern embedded applications exposes more security flaws~\cite{Ravi:2004}. 
 It is not an easy task to retrofit 
embedded systems with security mechanisms that were developed for more general purpose scenarios since the former 
\ca have constraints in processing power, memory, battery life, {\em etc.} and 
\cb are required to meet stringent requirements such as timing constraints.

%\begin{figure}[bht]
%\centering
%\includegraphics[width=1\columnwidth]{Figures/example.pdf}
%\caption{The sequence of system calls made when uploading an image through FTP and an example database of legitimate subsequences (length of 4). The malicious code makes the exact same sequence of the normal image uploading routine and hence sequence-based approaches cannot detect the malicious code.}
%\label{fig:example}
%\end{figure}

%Real-time embedded systems are increasingly attracting attackers looking to compromise their
%safety and security. Protecting such systems from the attacks is challenging due to the 
%ever-growing complexity of modern real-time embedded systems/applications; the additional 
%complexity, in turn, exposes more security flaws~\cite{Ravi:2004}. Thus, instead of attempting
%to prevent every possible security breach, we intend to detect intrusions by monitoring the 
%{\em behavior} of the application; hence, deviation from expected behavior is considered malicious.
%such as malicious code execution, by monitoring the system/application's \emph{behavior} has drawn
%a great deal of attention due to its ability to detect novel attacks; an anomaly, i.e., deviation, 
%from the expected/normal behavior is considered malicious~\cite{Denning:1987,Chandola:2009}. 
%Real-time embedded applications are a good fit for these types of security mechanisms due to the
%regularity in their execution behavior; the set of what constitutes legitimate behavior is 
%limited by design and also due to the (typically) small input set space.

Traditional behavior-based intrusion detection systems (IDS)~\cite{Denning:1987} rely on specific \emph{signals} such as
network traffic \cite{Handley:2001,Sommer10outsidethe}, 
control flow~\cite{Abadi:2009,Criswell:2014},
system calls \cite{Forrest:1996,Mutz:2006,Burguera:2011}, \textit{etc.} The use of system calls, especially in the form of sequences~\cite{Forrest:1996,Hofmeyr:1998,Warrender99detectingintrusion,Marceau:2000,Eskin01modelingsystem,Sun:2006}, has been extensively studied in behavior-based IDSes for general purpose systems since many malicious activities often use system calls to execute privileged operations on system resources. Because server, desktop and mobile applications exhibit rich, wildly varying behaviors across executions, such IDSes need to rely either (a) on complex models of normal behavior, which are expensive to run and thus unsuitable for an embedded system, or (b) on simple, partial models, which validate only small windows of the application execution at a time. This opens the door for attacks where variations of a valid execution sequence are replayed with slightly different parameters to achieve a malicious goal; on the other hand, the application would not execute that sequence of operations in a normal manner, every time.

\begin{figure}[t]
\centering
\includegraphics[width=0.8\columnwidth]{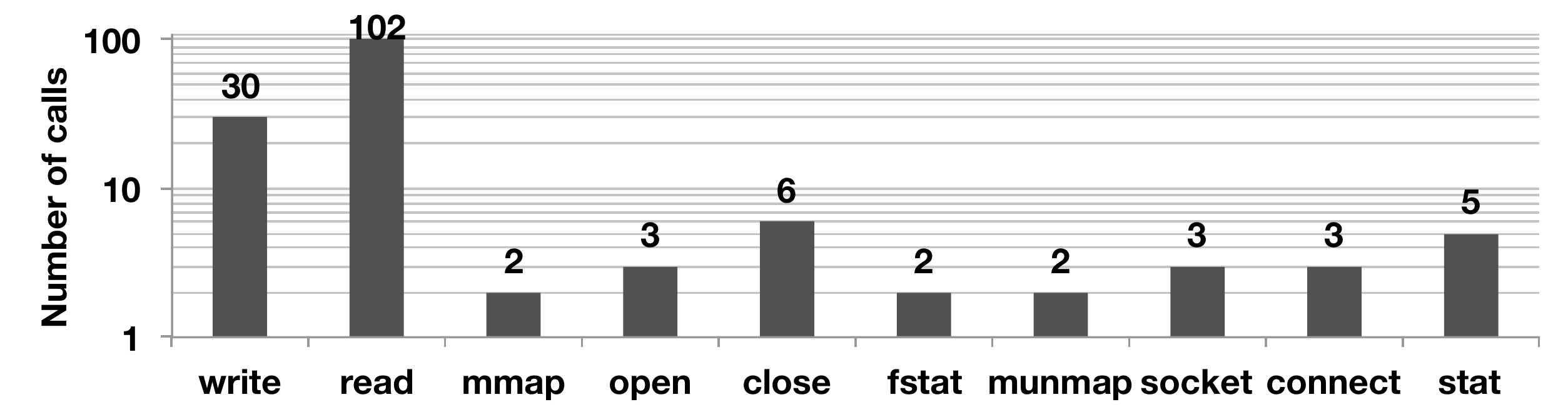}
\caption{An example system call frequency distribution (SCFD).}
\label{fig:syscallfreqdist}
\end{figure}

We observe that the very properties of embedded systems also make them amenable 
to the use of certain security mechanisms. The \emph{regularity} in their execution patterns means that we can 
detect intrusions by monitoring the {\em behavior} of such applications \cite{embeddedsecurity:s3a2012,yoonsecurecore2013,Zadeh:2014,YoonMHM:2015}; deviations from expected behavior can be considered to be malicious since the set of what constitutes legitimate behavior is often limited by design. %In particular the number of times an operation is invoked is often limited to a small range of values.
In this paper we present an intrusion detection mechanism for embedded systems using a {\em system call frequency distribution} (SCFD). Figure \ref{fig:syscallfreqdist} presents an example. It represents the {\em numbers of occurrences of each system call type} for each execution run of an application. The key idea is that the normal executions of an application whose behavior is predictable can be modeled by a small set of distinct system call distributions, each of which corresponds to a high-level \emph{execution context}. We use a {\em cluster analysis} to learn \emph{distinct} execution contexts from a set of SCFDs and to detect anomalous behavior using a {\em similarity metric} explained in Section~\ref{sec::analysis}.

Our detection method is lightweight, has a \emph{deterministic} time complexity -- hence, it fits well for resource-constrained embedded systems. This is due to the coarse-grained and concise representation of SCFDs. Although it can be implemented either at the operating system layer~\cite{Parmer:2007} or even used for offline analysis, we demonstrate an implementation 
on the SecureCore architecture~\cite{yoonsecurecore2013} that increases security. In Section~\ref{sec::architecture}, we show that minor modifications to a modern multicore processor enables us to monitor and analyze the run-time system call usage of applications in a secure, non-intrusive manner.

We implemented our prototype on Simics, a full-system simulator~\cite{SimicsPaper}. Due to the inherent limitations in simulation environment, we developed a proof-of-concept implementation based on an example embedded application and various attack scenarios that highlight the mechanism and benefits of the SCFD-based intrusion detection and also its limitations. The experimental results show that SCFDs can effectively detect certain types of abnormal execution contexts that are difficult for traditional sequence-based approaches~\cite{Chandola:2012,Sun:2006}. Detailed results including a comparison with an existing sequence-based technique is presented in Sections~\ref{sec::evaluation} and ~\ref{sec::eval_results}.

Hence, the high level contributions of this paper are:
\begin{enumerate}
	\item we introduce a lightweight method, utilizing the predictable nature of embedded system behaviors, with a deterministic time complexity for detecting anomalous execution contexts of embedded systems based on the {\em distribution of system call frequencies} (Section \ref{sec::analysis});
   %\item we introduce a lightweight method with a deterministic time complexity for detecting anomalous execution contexts of embedded systems based on the {\em distribution of system call frequencies} (Section \ref{sec::analysis});
	\item we present an architectural framework based on the SecureCore architecture for secure, non-intrusive monitoring and analysis of SCFDs (Section~\ref{sec::architecture});
	\item we demonstrate our techniques on a prototype implementation and evaluate its advantages and limitations using various attack scenarios that include a real attack (Sections~\ref{sec::evaluation} and \ref{sec::eval_results}).
\end{enumerate}

%% file: overview.tex
\section{Overview}
\label{sec:overview}

The main idea behind SCFD is to learn the normal system call profiles, \ie {\em patterns in system call frequency distributions}, collected during legitimate executions of a sanitized system. %Once the system is deployed, we observe the system call usage of the applications at run-time. If the run-time SCFD behavior deviates from profiles obtained during the analyses of the normal executions, then we claim that the application/system has been infected.
Analyzing profiles is challenging especially when such profiles change, often dramatically, depending on the execution modes and inputs. We address this issue by \emph{clustering} the distribution of system calls capturing legitimate behavior. 
Each cluster then can be a \emph{signature} that represents a high-level execution context, either in a specific mode or for similar input data. Then, given an observation at run-time,
we test how similar it is to each previously calculated cluster. If there is no strong statistical evidence that it is a result of a specific execution context then we consider the execution to be malicious with respect to the learned model.

\begin{figure*}[t]
\centering
\includegraphics[width=1\textwidth]{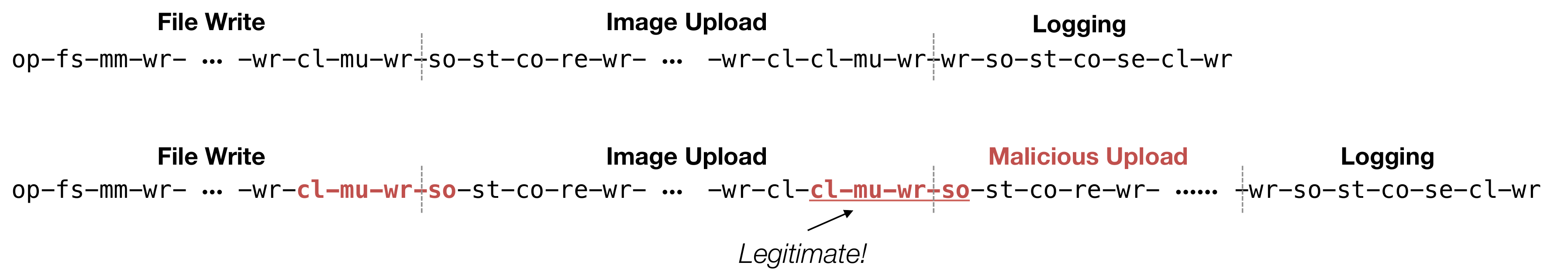}
\caption{Sample sequence of system calls made by the target embedded application used in the evaluation (Section~\ref{sec::evaluation}). The top sequence is from a normal execution. The bottom one is when a malicious code uses the exact same routine used by the normal execution for image uploading. The transition (that is underlined) is legitimate with respect to the set of legitimate sequences, and thus a sequence-based approach may fail to detect the malicious upload.\protect\footnotemark}
\label{fig:example_new}
%\begin{tabnote}%
%\Note{Abbr.:}{\textbf{ac}cess, \textbf{cl}ose, \textbf{co}nnect, \textbf{ex}ecve, \textbf{fs}tat, \textbf{ge}tuid, \textbf{mm}ap, \textbf{mu}nmap, \textbf{op}en, \textbf{re}ad, \textbf{se}ndto, \textbf{so}cket, \textbf{st}at, \textbf{wr}ite}
%\end{tabnote}%
\end{figure*}
\footnotetext{Abbreviations: \textbf{ac}cess, \textbf{cl}ose, \textbf{co}nnect, \textbf{ex}ecve, \textbf{fs}tat, \textbf{ge}tuid, \textbf{mm}ap, \textbf{mu}nmap, \textbf{op}en, \textbf{re}ad, \textbf{se}ndto, \textbf{so}cket, \textbf{st}at, \textbf{wr}ite}
  
\vspace{\baselineskip}
\noindent \textbf{Attacks against sequence-based IDSes: }Although sequence-based methods can capture detailed, temporal relations in system call usages, they may fail to detect abnormal \emph{execution contexts}.
This is because most sequence-based approaches rather profile the \emph{local, temporal} relations among system calls within a limited time frame. Figure~\ref{fig:example_new} highlights such a case. The sequence at the top is obtained from a part of normal execution of the target embedded application used in our evaluation (Section~\ref{sec::evaluation}). A smart attacker may use the very same routine to circumvent the detection process and upload the image to its own server right after the normal image uploading operation completes, as shown at the bottom. A sequence-based method may not detect this malicious activity if the model parameter (\eg the sequence length or the Markovian order~\cite{Chandola:2012}) is not carefully chosen, since the transition \texttt{cl-mu-wr-so} is \emph{not} abnormal with respect to what can be observed during normal executions.

In contrast to sequence-based techniques, our SCFD method may fail to detect a small local variation in system call sequences. However, as we show in this paper, it can easily detect abnormal deviations in high-level, naturally variable execution contexts such as the one illustrated above (Figure \ref{fig:example_new}) since the SCFD significantly changes due to the malicious execution. Also, if the attacker corrupts the integrity of the data (for instance, replaces the input or changes its size to downgrade its quality) then our method is able to detect such problems -- this is not easy for sequence-based methods as we explain in Section~\ref{subsec::eval_comparison}. Hence, by using these two approaches together, one can improve the overall accuracy of the system call-based IDS for embedded systems.

%The application(s) we monitor, however, may exhibit multiple normal execution contexts due to different operating modes and/or inputs. Representing such variations by a single behavioral context can lead to inaccuracies in the model due to the smoothing out of irregularities. Hence, we use a {\em cluster analysis} to identify \emph{distinct} execution contexts from a set of SCFDs and to detect anomalous behavior using a {\em similarity metric}, which we detail in Section~\ref{sec::analysis}.

%\subsection{Assumptions and Adversary Model}
%\label{subsec::assumptions}

\vspace{\baselineskip}
\noindent \textbf{Assumptions and Adversary Model: }The following assumptions are made in this paper: 
\begin{enumerate}
\item We consider an embedded application that executes in a \emph{repetitive fashion} -- We monitor and perform a legitimacy test at the end of each invocation of a task.
\item We limit ourselves to applications of which most of the possible execution contexts can be profiled ahead of time. Hence, the behavior model is learned under the \emph{stationarity} assumption -- this is a general requirement of most behavior-based IDS. 
%We limit ourselves to applications of which most of the possible execution contexts can be profiled ahead of time -- this is a general requirement of most behavior-based IDS. 
This can be justified by the fact that most embedded applications have a limited set of execution modes and input data 
fall within fairly narrow ranges. Also, a significant amount of analysis of embedded systems is carried out post-design/implementation anyways for a variety of reasons (to guarantee
predictable behavior for instance). Hence, the information about the usage of system calls can be rolled into such a-priori analysis. Our method may not work well for applications that do not exhibit execution regularities (\eg due to frequent user interactions).
\item The initial state of the application is trustworthy. The profiling is carried out prior to system deployment. Also, any updates to the applications must be accompanied by a repeat of the profiling process. The application(s) could be compromised after the profiling 
stage, but we assume that the stored profile(s) cannot be tampered with. Again, such 
(repeat) analysis is typical in such systems -- \eg anytime the system receives updates, including changes to the operating system or the processor architecture.
\item We consider threat models that involve changes to the behavior of system 
call usage. If an attack does not invoke or change any system calls, the 
activity at least has to affect executions afterward so that the future system call 
usage may change. The methods in this paper, as they stand, cannot detect attacks 
that never alter system call usage and that just replace certain system calls by 
\emph{hijacking} them (\eg altering kernel system call table) \cite{syscall_hijacking}.  
\item We consider malicious code that can be secretly embedded in the application,
either by remote attacks or during upgrades. The malicious code activates itself at some
point {\em after} system initialization. We are not directly concerned with how the 
malicious code gained entry, but focus more on what happens after that.
\end{enumerate}

As mentioned above, we assume that malware will exhibit a different pattern of system call usage. 
For example, malware that leaks out a sensitive information would make use of network-related 
system calls (\eg \texttt{socket}, \texttt{connect}, \texttt{write}, \etc) thus changing the
frequencies of these calls. %Also, changes in control flow due to, \eg \emph{shellcode} execution~\cite{AlephOne} could result in changes to the frequencies of the calls. 

%\begin{comment}
%\subsection{Attacker Model}
%\label{subsec::attacker_model}
%
%Thus, in this paper, we present an intrusion detection method for real-time embedded applications using a {\em system call frequency distribution} (SCFD), \ie a vector of non-negative integers, where each entry
%represents \emph{the number of occurrences of a particular system call type} for an execution. We treat the application under monitoring as a black-box and count how many times each system call type has been used. Figure 
%\ref{fig:syscallfreqdist} shows an example SCFD obtained from the application used in Section \ref{sec::evaluation}. The underlying idea is that normal executions would follow an expected pattern of system call usage -- this assumption holds for real-time systems that exhibit very little variations during execution. Another, related, assumption is that malware will exhibit a different pattern of system call usage. For example, malware that leaks out a sensitive information would use some network-related system calls (\eg \texttt{socket}, \texttt{connect}, \texttt{write}, \etc) thus changing the frequencies of these calls. Also, changes in control flow due to, for example, \todo{modified}a \emph{shellcode} execution~\cite{AlephOne} or a \emph{return-to-libc} attack~\cite{Nergal:2001} could result in changes to the frequencies of the calls. 
%\end{comment}

%% file: Analysis.tex
\section{Intrusion Detection Using Execution Contexts Learned from System Call Distributions}
\label{sec::analysis}

We now present our novel methods to detect abnormal execution contexts in embedded applications by monitoring changes in system call frequency distributions.

\subsection{Definitions}
\label{subsec::definitions}

Let $\mathcal{S}=\{s_1, s_2, \ldots, s_{D}\}$ be the set of all system calls provided by an operating system, where $s_d$ represents the 
system call of type $d$. During the $n^{th}$ execution of an application, it calls a multiset $\sigma^n$ of $\mathcal{S}$. Let us denote 
the $n^{th}$ \emph{system call frequency distribution} (or just \emph{system call distribution}) as 
$\mathbf{x}^n = [m(\sigma^n, s_1), m(\sigma^n, s_2), \ldots, m(\sigma^n, s_{D})]^T$, where $m(\sigma^n, s_d)$ is the multiplicity of 
the system call of type $d$ in $\sigma^n$.
Hereafter, we simplify $m(\sigma^n, s_d)$ as $x^n_d$. Thus, $\mathbf{x}^n = [x^n_1, x^n_2, \ldots, x^n_D]^T$.

We define a \emph{training set}, \ie the execution profiles of a sanitized system, as a set of $N$ system call frequency distributions collected from $N$ executions, and is denoted by $\mathbf{X} = [\mathbf{x}^1, \mathbf{x}^2, \ldots, \mathbf{x}^N ]^T$. The clustering algorithm (Section~\ref{sec::global_k_means}) then maps each $\mathbf{x}^n\in \mathbb{N}^D$ to a cluster $c_i\in \mathcal{C}=\{c_1, c_2, \ldots, c_k\}$. We denote by $c: \{\mathbf{x}^1,\cdots, \mathbf{x}^N\} \rightarrow \mathcal{C}$ the cluster that $\mathbf{x}^n\in \mathbf{X}$ belongs to.
%$c(\mathbf{x}^n)=c_i$ if $\mathbf{x}^n$ belongs to cluster $c_i$. 

%Our algorithm uses the Mahalanobis \emph{distance} metric \cite{Mahalanobis1936}, which will be explained in Section~\ref{sec::single}, to measure how similar $\mathbf{x}^n$ to each $c_i$. We denote it as $dist(\mathbf{x}^n, c_i)$.\\

\subsection{Learning a Single Execution Context}
\label{sec::single}

The variations in the usage of system calls will be limited if the application under monitoring has a simple execution context. In such a case, it is reasonable to consider that the executions follow a certain distribution of system call frequencies, clustered around a \emph{centroid}, and cause a small variation from it due to, for example, input data or execution flow. This is a valid model for many embedded systems since the code in such system tends to be fairly limited in what it can do. Hence, such analysis is quite powerful in detecting variations and thus catching intrusions.

%\begin{figure}[htb]
\begin{figure}[t]
\centering
\includegraphics[width=0.65\columnwidth]{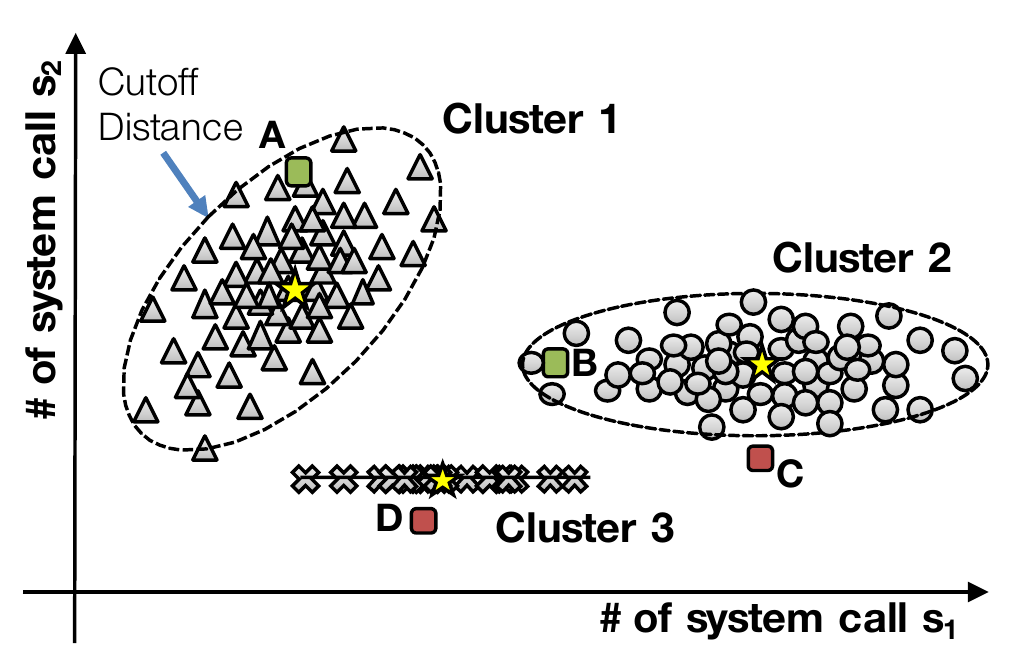}
\caption{System call frequency distributions for $\mathcal{S}=\{s_1, s_2\}$ and clusters. The gray-colored objects are SCFDs in the training set. Each star-shaped point inside each cluster is its centroid. The ellipsoid around each cluster draws the cutoff line of the cluster; the points inside of the line are legitimate with respect to the cluster.}
\label{fig:kmeans}
\end{figure}

For a multivariate distribution, the mean vector $\boldmath\mu =$ $[\mu_1$, $\mu_2$, $\ldots$ ,$\mu_D]^T$, where $\mu_d = (\sum_{n}^{N} x_d^n)/N$,
can be used as the centroid. Figure~\ref{fig:kmeans} plots the frequency distributions of two system call types
(\ie $D=2$). For now, let us consider only the data points (triangles) on the left-hand side of the graph. The data points are clustered 
around the star-shaped marker that indicates the centroid of the distribution formed by the points. Now, given a new observation from the 
monitoring phase, \eg the point marked `\textsf{A}', a {\em legitimacy test} can be devised that tests {\em the likelihood} that such an
observation is actually part of the expected execution context. %This can be done by calculating the distance between the star marker and `\textsf{A}', \ie \emph{how far} the new observation is from the centroid. 
This can be done by measuring \emph{how far} the new observation is from the centroid. 
Here, the key consideration is on the \emph{distance} measure for testing legitimacy. 

One may use the Euclidean distance between the new observation $\mathbf{x}^*$ and the mean vector of a cluster, \ie $|| \mathbf{x}^* - \boldmath\mu|| = \sqrt{ (\mathbf{x}^*-\boldmath\mu)^T  (\mathbf{x}^*-\boldmath\mu)}$. Although the Euclidean distance (or $L^2$-norm) is simple and straightforward to use, the distance is built on a strong assumption that each coordinate (dimension) contributes \emph{equally} while computing the distance. In other words, the same amount of differences in $x_1^n$
and $x_2^n$ are considered equivalent even if, \eg a small variation in the usage of system call $s_2$ is the stronger indicator of abnormality than system call $s_1$. Thus, it is more desirable to allow such a variable contribute more. For this reason, we use the 
\emph{Mahalanobis} distance \cite{Mahalanobis1936}, defined as:
\begin{align}
dist_M (\mathbf{x}^n, \mathbf{X}) = \sqrt{ (\mathbf{x}^n-\boldmath\mu)^T \boldmath\Sigma^{-1} (\mathbf{x}^n-\boldmath\mu) },\nonumber
\end{align}
for a group of data set $\mathbf{X}$, where $\boldmath\Sigma$ is the covariance matrix of $\mathbf{X}$.\footnote{$\boldmath\Sigma$ is the positive definite. If we set $\boldmath\Sigma = \mathbf{I}$, the Mahalanobis distance is equivalent to the Euclidean distance. Thus, the Mahalanobis 
distance is more expressive than the Euclidean distance.} Notice that the existence of $\boldmath\Sigma^{-1}$ is the necessary condition to define the Mahalanobis distance;
 \ie the difference of the frequency of each system call from the mean (\ie what is expected) is augmented by the \emph{inverse of its variance}. 

Accordingly, if we observe a small variance for certain system calls during the training, \eg \texttt{execve} or \texttt{socket}, 
we would expect to see a similar, small, variation in the usage of the system calls during actual executions as well. On the other hand, 
if the variance of a certain system call type is large, \eg \texttt{read} or \texttt{write}, the Mahalanobis distance metric gives a 
small weight to it in order to keep the distance (\ie abnormality) less sensitive to changes in such system calls. Cluster $2$ in Figure~\ref{fig:kmeans} shows an example of the advantage of 
using the Mahalanobis distance over the Euclidean distance. Although \textsf{C} is closer to the centroid than \textsf{B} is in terms of the 
Euclidean distance, it is more reasonable to determine that \textsf{C} is an outlier and \textsf{B} is legitimate because we have not seen (during the normal executions) frequency distributions such as the one exhibited by \textsf{C} while we have seen a statistically meaningful amount
of examples like \textsf{B}. As an extreme case, let us consider \textsf{D} which is quite close to Cluster $3$'s center in terms of the Euclidean distance. However, it should be considered malicious because $s_2$ (i.e. the $y$-axis) should never vary. 

Using covariance values also make it possible to learn \emph{dependencies} among different system call types. For instance, an occurrence of the 
\texttt{socket} call usually accompanies \texttt{open} and many \texttt{read} or \texttt{write} calls.  Thus, we can easily expect that changes in \texttt{socket}'s frequency would also lead to variations in the frequencies of \texttt{open}, \texttt{read} and 
\texttt{write}. Cluster $1$ in Figure~\ref{fig:kmeans} is such an example that shows covariance between the two system call types. 
On the other hand, they are independent in Cluster $2$ and $3$. Thus, using the Mahalanobis distance we can not only learn how 
many occurrences of each individual system call should exist but also how they should vary together.

Now, given a set of system call distributions, $\mathbf{X} = [\mathbf{x}^1, \mathbf{x}^2, \ldots, \mathbf{x}^N ]^T$, we calculate the mean vector, $\boldmath\mu$, and the covariance matrix, $\boldmath\Sigma$, for this data set. It then can be represented as a single cluster, $c$, whose centroid is defined as $(\boldmath\mu, \boldmath\Sigma)$. Now, the Mahalanobis distance of a newly observed SCFD, $\mathbf{x}^*$, from the centroid is
\begin{align}\label{eq:mahal2}
dist(\mathbf{x}^*, c) = \sqrt{ (\mathbf{x}^*-\boldmath\mu)^T \boldmath\Sigma^{-1} (\mathbf{x}^*-\boldmath\mu) }.
\end{align}
If this distance is greater than a cutoff distance $\theta$, we consider that the execution to be malicious. For example, \textsf{B} in Figure~\ref{fig:kmeans} is considered legitimate w.r.t. Cluster $2$. One analytic way to derive this threshold, $\theta$, is to think of the Mahalanobis distance w.r.t. the multinomial normal distribution, 
\begin{align}
%p(\mathbf{x}^*) = \frac{1}{\sqrt{|\boldmath\Sigma|(2\pi)^D} } \exp\Big( -\frac{1}{2} dist(\mathbf{x}^*, c)^2\Big).
p(\mathbf{x}^*) = {\sqrt{|\boldmath\Sigma|(2\pi)^D}}^{-1} \exp\Big( -\frac{1}{2} dist(\mathbf{x}^*, c)^2\Big).
\end{align}
That is, we can choose a $\theta$ such that the p-value under the null hypothesis is less than a significant level $p_0$, \eg
$1\%$ or $5\%$. Appendix \ref{appendix::cutoff} explains how to calculate $\theta$ given a $p_0$.

\subsection{Learning Multiple Execution Contexts Using Global k-means}
\label{sec::global_k_means}

In general, an application may show widely varying system call distributions due to multiple execution modes and varying inputs. In such scenarios, finding a single cluster/centroid for the whole set can result in inaccurate models because it would include many non-legitimate points that belong to none of the execution contexts -- \ie the empty space between clusters in Figure~\ref{fig:kmeans}. Thus, it is more desirable to consider that observations are generated from a set of \emph{distinct} distributions, 
each of which corresponds to one or more execution contexts. Then, the legitimacy test for a new observation $\mathbf{x}^*$ is 
reduced to identifying the \emph{most probable} cluster that may have generated $\mathbf{x}^*$. 
If there is no strong evidence that $\mathbf{x}^*$ is a result of an execution corresponding to any cluster
then we determine that $\mathbf{x}^*$ is most likely due to malicious execution.  

Suppose we collect a training set $\mathbf{X} = [\mathbf{x}^1, \mathbf{x}^2, \ldots, \mathbf{x}^N ]^T$ where 
$\mathbf{x}^n \in \mathbb{N}^D$. To learn the distinct distributions, we use the $k$-means algorithm \cite{Lloyd82} to partition the $N$ data points on a
$D$-dimensional space into $k$ clusters. The $k$-means algorithm works as follows:
\begin{enumerate}
\item Initialization: Create $k$ initial clusters by picking $k$ random data points from $\mathbf{X}$.
\item Assignment: For each $\mathbf{x}^n \in \mathbf{X}$, assign it to the closest cluster $c(\mathbf{x}^n)$, \ie  
\begin{align}
\label{eq:closest_cluster}
c(\mathbf{x}^n) = \arg\min_{c_{k}\in \mathcal{C}} dist(\mathbf{x}^n, c_{k}).
\end{align}
\item Update: Re-compute the centroid (\ie $\boldmath\mu$ and $\boldmath\Sigma$) of each cluster based on the new assignments. 
\end{enumerate}
The algorithm repeats steps {\em (2)} and {\em (3)} until the assignments stop changing. Intuitively speaking, the algorithm 
keeps updating the $k$ centroids until the total distance of each point $\mathbf{x}^n$ to its cluster,
\begin{align}\label{eq:total_dist}
total\mbox{-}dist(\mathbf{X}, \mathcal{C}) = \sum_{n=1}^{N} dist( \mathbf{x}^n, c(\mathbf{x}^n) ),
\end{align}
is minimized.

\begin{algorithm}[tb]
{
\caption{\textsc{Global k-means}($\mathbf{X}, \mathtt{MAX_K}, \mathtt{Bound_{TD}}$)   } \label{alg:global-k-means}
\begin{algorithmic}[1]
\STATE \COMMENT {$\mathbf{X}$: the training set}
\STATE \COMMENT {$\mathtt{MAX_K}$: the maximum number of clusters}
\STATE \COMMENT {$\mathtt{Bound_{TD}}$: the total distance bound}

\STATE Create $c_1$ with $\mathbf{X}$. Calculate $\boldsymbol\mu_1$ and $\boldsymbol\Sigma_1$.
\STATE $\mathcal{C} \gets \{c_1\}$
\STATE $k \gets 2$
\STATE $Min_{TD} \gets \infty$
\WHILE {$k\le \mathtt{MAX_K}$ or $total\mbox{-}dist(\mathbf{X}, \mathcal{C})>\mathtt{Bound_{TD}}$}
	\FOR{$n=1,\ldots,N$}
		\STATE Create $c_{k}$ with $\mathbf{x}^n$ as its initial point. 
		\STATE $\mathcal{C}' \gets $\textsf{k-means}$(\mathbf{X}, \mathcal{C} \cup c_{k})$ 
		\IF { $total\mbox{-}dist(\mathbf{X}, \mathcal{C}') < Min_{TD}$ }
			\STATE \COMMENT {Note: The best clustering for $k$ so far}
			\STATE $\mathcal{C}^* \gets \mathcal{C}'$ 
			\STATE $Min_{TD} \gets total\mbox{-}dist(\mathbf{X}, \mathcal{C}')$
		\ENDIF
	\ENDFOR
	\STATE $\mathcal{C} \gets \mathcal{C}^*$
	\STATE $k \gets k + 1$
\ENDWHILE
\RETURN $\mathcal{C}$
\end{algorithmic}}
\end{algorithm}

The $k$-means algorithm requires a strong assumption that we already know $k$, the number of clusters. However, this assumption does not hold in reality because the number of distinct execution contexts is not known ahead of time. Moreover, the accuracy of the final model heavily depends on the initial clusters chosen randomly.\footnote{Finding the optimal assignment in the $k$-means algorithm with the Euclidean distance is NP-hard. Thus, finding the optimal assignments with the Mahalanobis distance is at least NP-hard because the Mahalanobis distance is more general than the Euclidean distance.} Hence, we use the \emph{global $k$-means} method \cite{Likas2003} to find the number of clusters as well as the initial assignments that lead to \emph{deterministic accuracy}. Algorithm~\ref{alg:global-k-means} illustrates the global $k$-means algorithm. Given a training set $\mathbf{X}$ of $N$ system call frequency distributions, the algorithm finds the best number of clusters and assignments. This is an incremental learning algorithm that starts from a single cluster, $c_1$, consisting of the entire data set. In the case of $k=2$, the algorithm considers each $\mathbf{x}^n\in \mathbf{X}$ as the initial point for $c_2$ and runs the assignment and updates steps of $k$-means algorithm. After $N$ trials, we select the final centroids that resulted in the smallest total distance calculated by Eq. \eqref{eq:total_dist}. These two centroids are then used as the initial points for the two clusters, respectively, in the case of $k=3$. This procedure repeats until either $k$ reaches a pre-defined $\mathtt{MAX_K}$, the maximum number of clusters, or the total distance value becomes less than the total distance bound $\mathtt{Bound_{TD}}$. Note that the total distance in Eq. \eqref{eq:total_dist} decreases monotonically with the number of clusters. For example, if every point is its own cluster then the total distance is zero since each point itself is the centroid. 

The original algorithm assumes the Euclidean distance. As explained above, we use the Mahalanobis distance as in Eq. \eqref{eq:mahal2}. Meanwhile, \textsf{k-means}$(\mathbf{X}, \mathcal{C})$ (line 11) is the standard $k$-means algorithm without the random initialization; it assigns the points in $\mathbf{X}$ to a $c_k \in \mathcal{C}$, update the centroids, repeats until stops, and then returns the clusters with the \emph{updated} centroids. The standard $k$-means algorithm uses the Euclidean distance and thus the centroids of the initial clusters are the data points that were picked first. Remember, however, that the Mahalanobis distance requires a covariance matrix. Since there would be 
only one data point in each initial cluster we use the global covariance matrix of the entire data set $\mathbf{X}$
for the initial clusters. After the first iteration, however, the covariance matrix of each cluster is updated using
the data points assigned to it.

%It is an incremental learning algorithm that starts from a single cluster consisting of the entire data set and stops when either $k$ reaches a pre-defined $\mathtt{MAX_K}$ or the total distance value becomes less than the total distance bound $\mathtt{Bound_{TD}}$. Appendix~\ref{appendix::global_k_means} explains how to apply the global k-means algorithm with the Mahalanobis distances.

\vspace{\baselineskip}
The clustering algorithm finally assigns each data point in the training set into a cluster. Then, each cluster $c_i\in \mathcal{C}$ 
can be represented by the centroid, $(\boldmath\mu_i, \boldmath\Sigma_i)$, that now makes it possible to calculate the 
Mahalanobis distance of a newly observed SCFD $\mathbf{x}^*$ to each cluster using Eq. \eqref{eq:mahal2}. The legitimacy test of 
$\mathbf{x}^*$ is then performed by finding the closest cluster, $c^*$, using Eq. \eqref{eq:closest_cluster}. Thus, if 
\[dist(\mathbf{x}^*, c^*)= \min_{c_{i}\in \mathcal{C}} dist(\mathbf{x}^*, c_{i})  > \theta\]
for a given threshold $\theta$, we determine that the execution does not fall into any of the execution contexts specified by 
the clusters since $dist(\mathbf{x}^*, c_i)  >\theta$ for all $i=1,\ldots,k$. We then consider the execution to be malicious. As an example, for the new observation \textsf{C} in 
Figure~\ref{fig:kmeans}, Cluster $2$ is the closest one and \textsf{C} is outside its cutoff distance. Thus, we consider that \textsf{C}
is malicious. Note that, as shown in the figure, the same cutoff distance defines different 
ellipsoids for different clusters; each ellipsoid is a equidistant line from the mean vector measured in terms of the \emph{Mahalanobis} distance. Thus, a cluster with small variances (\ie less varying execution context) would have a smaller ellipsoid in the Euclidean space.

\subsection{Dimensionality Reduction}
\label{subsec::dimension}

The number of system call types, \ie $D$, is quite large in general. Thus, the matrix calculations in Eq. \eqref{eq:mahal2} might result in an unacceptable amount of analysis overhead.\footnote{Note that $\boldmath\mu$ and $\boldmath\Sigma$ values are calculated 
from the clustering algorithm which is an \emph{offline} analysis. Thus we store $\boldmath\Sigma^{-1}$ for computational efficiency.} However, embedded applications normally use a limited subset of system calls. Furthermore, 
we can significantly reduce the dimensionality by ignoring system call types that never vary. Consider 
Cluster $3$ from Figure~\ref{fig:kmeans}. Here, $x_2$ can be ignored since we can reasonably expect it to never vary during the normal execution.\footnote{In fact, $s_2$ cannot be ignored in the example depicted in Figure~\ref{fig:kmeans} since its variance is non-zero in clusters $1$ and $2$.} Thus, before running the clustering algorithm, we reduce $\mathcal{S}$ to $\mathcal{S}' = \{s_{d_1}, s_{d_2}, \ldots, s_{D'}\}$, where $D' \le D$, such that the variance of $x_d$ for each $s_{d} \in \mathcal{S}'$ is non-zero in the entire training set $\mathbf{X}$. However, we should still be able to detect any changes in such system calls that never varied (including those that never appeared). Thus, we merge all such $x_d$ in $\mathcal{S} - \mathcal{S}'$; the sum should not change in normal executions. In case $D'$ is still large, one may apply a statistical dimensionality reduction technique such as Principal Component Analysis (PCA)~\cite{Jolliffe:2002} or Linear Discriminant Analysis (LDA)~\cite{fisher36lda}.

%(Longer version) Thus, before running the clustering algorithm presented above, we reduce $\mathcal{S}$ to $\mathcal{S}' = \{s_{d_1}, s_{d_2}, \ldots, s_{D'}\}$, where $D' \le D$, such that the variance of $x_d$ in the entire training set $\mathbf{X}$ is zero for each $s_{d} \in \mathcal{S} - \mathcal{S}'$. Note that $\mathcal{S} - \mathcal{S}'$ includes any system call types that never appeared in the training set. However, we should still be able to detect any changes in such system calls. Thus, we merge all the variables in $\mathcal{S} - \mathcal{S}'$ by calculating the sum $\sigma = \sum_{s_d \in \mathcal{S} - \mathcal{S}'} x_d$, where $x_d = x_d^1 = x_d^2 = \cdots = x_d^N$. Then, in the monitoring phase, we test if $\sigma - (\sum_{s_d \in \mathcal{S} - \mathcal{S}'} x_d^*)$ is non-zero. If it is, the application shows a variation in $\mathcal{S} - \mathcal{S}'$, which should be considered suspicious. In case $D'$ is still large, one may apply a statistical dimensionality reduction technique such as Principal Component Analysis (PCA)~\cite{Jolliffe:2002} or Linear Discriminant Analysis (LDA)~\cite{fisher36lda}.

%% file: architecture.tex
\section{Architectural Support for SCFD Monitoring}
\label{sec::architecture}

Most the existing system call-based intrusion detection systems rely 
on the operating system to provide the information, say, by use of auditing modules  \cite{snare,Parmer:2007}. 
While this provides the ability to monitor extensive resources such as system call arguments,
it requires the operating system itself to be trustworthy. In this paper we 
avoid this problem by proposing a new architectural framework that requires minor micro-architecture modifications. The architecture builds 
upon the SecureCore architecture~\cite{yoonsecurecore2013} that enables a trusted on-chip entity, \eg a \emph{secure core}, to continuously monitor the run-time behavior of applications on another, potentially untrusted entity, the \emph{monitored core}, {\em in a non-intrusive manner}. 
In this section we describe the modifications to SecureCore that enable us to monitor system calls.  
We refer interested readers to \cite{yoonsecurecore2013} for the full details about the SecureCore architecture.

\subsection{Overview}
\label{subsec::overview}

%\begin{figure}[htb]
\begin{figure}[t]
\centering
\includegraphics[width=0.55\columnwidth]{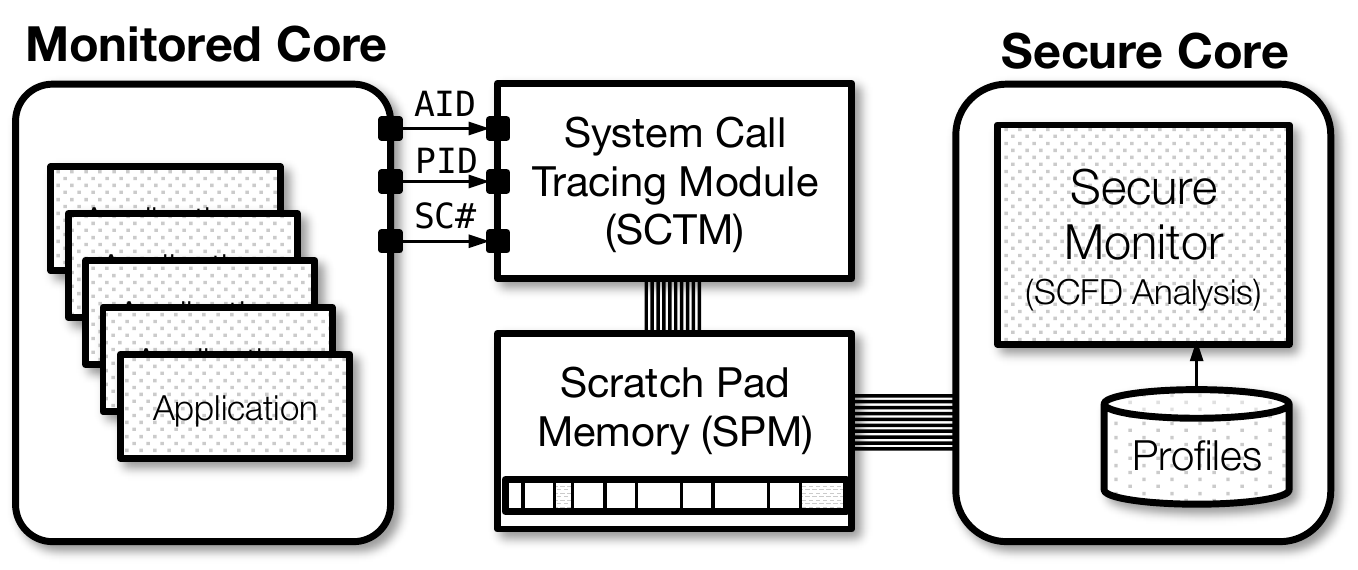}
\caption{The SecureCore architecture for SCFD monitoring. The SCTM traces the system calls made by applications running on the monitored core. The traces, written in the SPM, are retrieved by the secure monitor to perform legitimacy test for the executions.}
\label{fig:securecore}
\end{figure}

Figure~\ref{fig:securecore} shows the overall architecture for system call monitoring.
It consists of \ca a secure core, \cb a monitored core, \cc an on-chip system call tracing module (SCTM) and \cd a scratch pad memory (SPM). The secure core uses the SCTM and SPM to monitor the usage of system calls by applications executing on the monitored core. The SCTM extracts relevant information from the monitored core and then writes it to the SPM. A monitoring process on the secure core then uses this information to check whether the run-time behavior has deviated from the expected behavior that we profiled using the method described in Section~\ref{sec::analysis}. 

Note: We capture the profile of \emph{normal} executions in a similar manner: the monitoring process collects SCFDs using the SCTM and SPM under trusted conditions. We then apply the learning algorithm from Section~\ref{sec::analysis}. The resulting normal profile (one per application) is then stored in a secure memory location.

\subsection{System Call Tracing Module (SCTM)}
\label{sec::sctm}

The system call tracing module (SCTM) tracks how many times each application on the monitored core uses each system call type (\ie SCFDs). The main point is to catch the moment each call is invoked. We are able to do this because, in most processor architectures, a specific instruction is designated for this very purpose, \ie for triggering system calls. The calling conventions vary across processor architectures and operating systems. In the PowerPC architecture, that our prototype is based on, an \texttt{sc} instruction issues a system call based on a number stored in the \texttt{r0} register~\cite{e600}. The actual call is then handled by the operating system kernel. Hence, the execution of the {\tt sc} instruction denotes the invocation of a system call.

\begin{figure}[t]
\centering
\includegraphics[width=0.7\columnwidth]{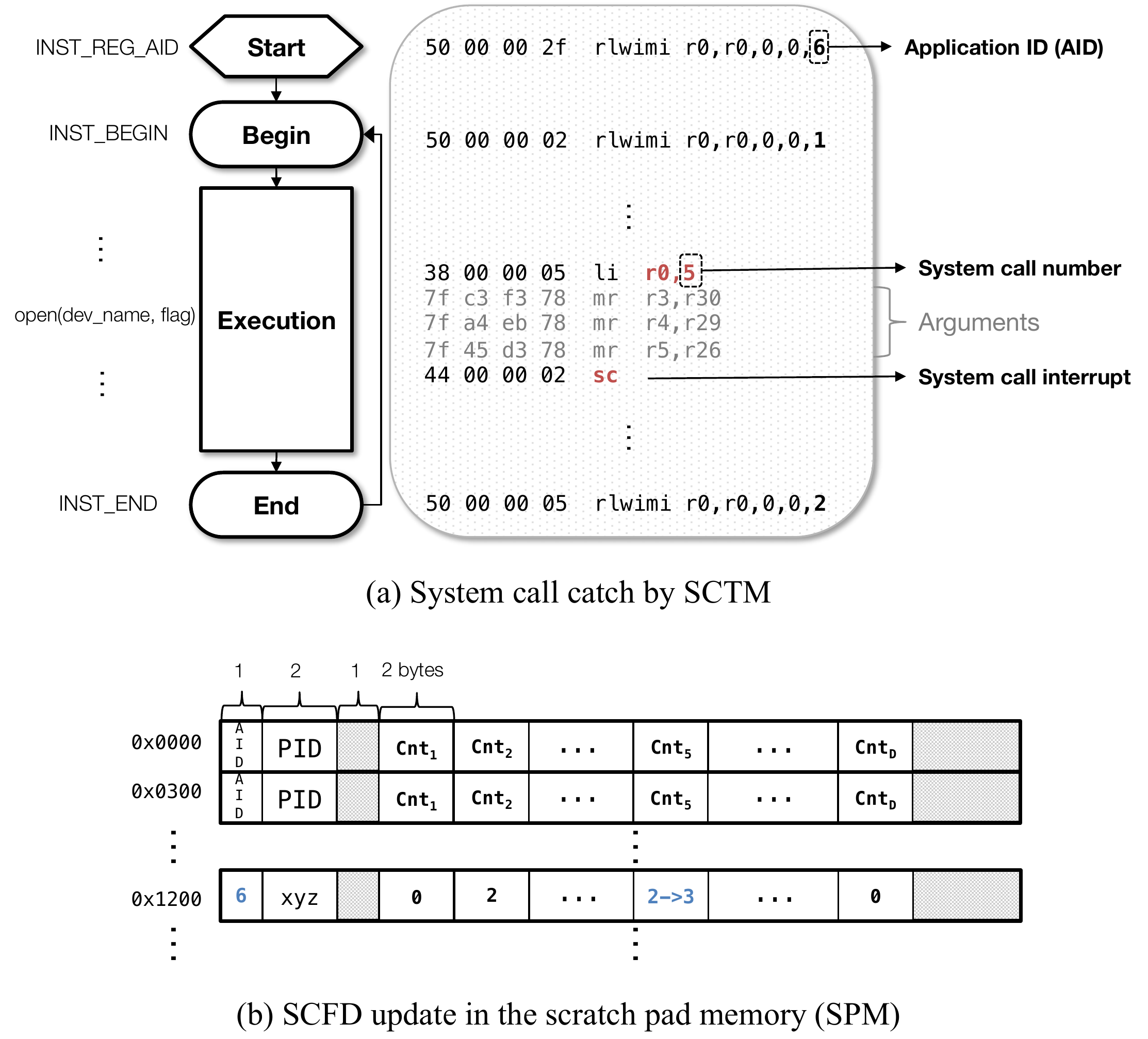}\caption{(a) The System Call Tracing Module (SCTM) catches system call executions by looking at the instruction decoder. (b) Upon execution of a system call, the corresponding counter (e.g., Cnt5 for \texttt{open}) of the SCFD entry mapped to the application is incremented. The gray areas are for alignment padding. }
\label{fig:instruction_and_spm}
\end{figure}

Another piece of information that we require is {\em who} initiated the call; hence we introduce a new instruction to help with identifying the requester. Figure~\ref{fig:instruction_and_spm}(a) describes the process by which the SCTM gathers the required information from an application. When an application starts, it registers its \emph{Application ID} (AID) and \emph{Process ID} with the SCTM. Here, the AID is a unique numerical value assigned to each application. It is used to let the secure monitor (on the secure core) locate the correct profile. Once the registration is complete, the secure monitor is able to map each application to corresponding PIDs. The above registration process is carried out by a special instruction, \texttt{INST\_REG\_AID}, as described in the figure.\footnote{\texttt{rlwimi} instruction is the \emph{Rotate Left Word Immediate Then Mask Insert} instruction in the PowerPC ISA~\cite{e600}. An execution of \texttt{rlwimi 0,0,0,0,i} for $0 \le i \le 31$ is equivalent to a nop -- hence, we used it for our purposes.} The special instruction has other modes as well: \ci \texttt{INST\_BEGIN} and \cii \texttt{INST\_END} that demarcate the region where we are tracking the usage of system calls. Once \texttt{INST\_END} completes, the monitor retrieves the data collected from the recently completed region of execution and applies the detection algorithm. The data is reset with the execution of an \texttt{INST\_BEGIN}. While an attacker may try to execute malicious code block before \texttt{BEGIN} or after \texttt{END} to avoid detection, we can catch such situations
because there should be no system call execution during that point in the code. Thus, in such cases, the SCTM would immediately raise an alarm. Also, an attack may skip some of all of the special instructions or modify any of the values. 
Again, these cannot help the attacker hide malicious code execution because the system call  distribution would need 
to be consistent with the profile. Also, watchdog timers can be used to check whether the applications are executing 
the special instructions in time.

% %\begin{figure}[htb]
% \begin{figure}[t]
% \centering\centering
% \includegraphics[width=0.93\columnwidth]{Figures/spm.pdf}
% \caption{The layout of the scratch pad memory. Each entry stores a system call frequency distribution of the application specified by PID. The gray areas are for alignment padding. }
% \label{fig:spm}
% \end{figure} 

Figure~\ref{fig:instruction_and_spm}(a) shows how an \texttt{open} system call invocation is detected. As explained above, the system call number, $5$ in Linux, is written to the \texttt{r0} register. Thus, by looking at the value of the register, we can track which system call is being invoked. When an $\texttt{sc}$ instruction is executed, SCTM takes the PID register and the \texttt{r0} register values. It then updates the corresponding \emph{SCFD entry} in the scratch pad memory (see Figure~\ref{fig:instruction_and_spm}(b)). An entry is a contiguous memory region of length $2D+4$ bytes, where $D$ is the total number of system calls provided by the OS. Using the PID, SCTM locates the corresponding entry and then increments the \emph{counter} of system call \texttt{d} if the value in \texttt{r0} was \textit{d}. The sizes of the SPM and each entry field are implementation-dependent. In our implementation, we assume at most 382 system call types (that are enough to cover most Linux implementations) that results in the size of an entry being 768 bytes at most. Thus, an SPM of size 8KB can provide a space for around $10$ applications to be monitored simultaneously. The SPM can be accessed only by the secure core. %It is mapped to a range of the secure core's address space that is protected by a hypervisor~\cite{yoonsecurecore2013} or a hardware-enforced memory protection mechanism~\cite{ARMTrustZoneNew}. 
When an \texttt{INST\_END} is executed, the secure monitor reads the corresponding entry from the SPM,\footnote{An interrupt can be raised by SCTM to inform the secure monitor of the execution of \texttt{INST\_END}. However, if possible, it should be avoided because the secure core can be continuously interrupted by a compromised application on the monitored core thus degrading its ability to perform the monitoring. If necessary, the SCTM should be configured to block consecutive \texttt{INST\_END} instructions. Of course, it is more preferable for the monitor to poll the SCTM.} a new SCFD, finds the profile for the corresponding application using the PID to AID map, then verifies the legitimacy of the SCFD.

%The secure monitor then verifies the legitimacy of the execution by using \ca the observed system call frequency distribution in the entry, \ie $\mathbf{x}^*$, and \cb the previously calculated profile, \ie the clusters' centroids, obtained when the system was in a known good state.

%% file: Evaluation.tex
\section{Evaluation Framework}
\label{sec::evaluation}

In this section, we first present the implementation details for our prototype (Section \ref{subsec::implementation}), the application model for our experiments (Section \ref{subsec::application_model}) and some attack scenarios that are relevant to this application (Section \ref{sec::attack_model}).

%In this section, we first present the implementation details for our prototype (Sec. \ref{subsec::implementation}), the application model for our experiments (Sec. \ref{subsec::application_model}), some attack scenarios that are relevant to this application (Sec. \ref{sec::attack_model}). We then evaluate the efficiency and the effectiveness of the proposed approach, by comparing it with one of the well-known sequence-based approach (Sec. \ref{sec::eval_eval}, \ref{subsec::overhead} and \ref{subsec::eval_comparison}). And finally we discuss the limitation and possible improvements (Sec. \ref{subsec::limitations}).

%In this section, we present a prototype of our intrusion detection method implemented on a full-system simulator. 

%In this section, we evaluate our intrusion detection method using a prototype implemented on a full-system simulator and also compare it with a sequence-based approach to show how SCFD-based method can complement the sequence-based approaches.

\subsection{System Implementation}
\label{subsec::implementation}

%\begin{figure}[htb]
\begin{figure}[t]
\centering
\includegraphics[width=0.8\columnwidth]{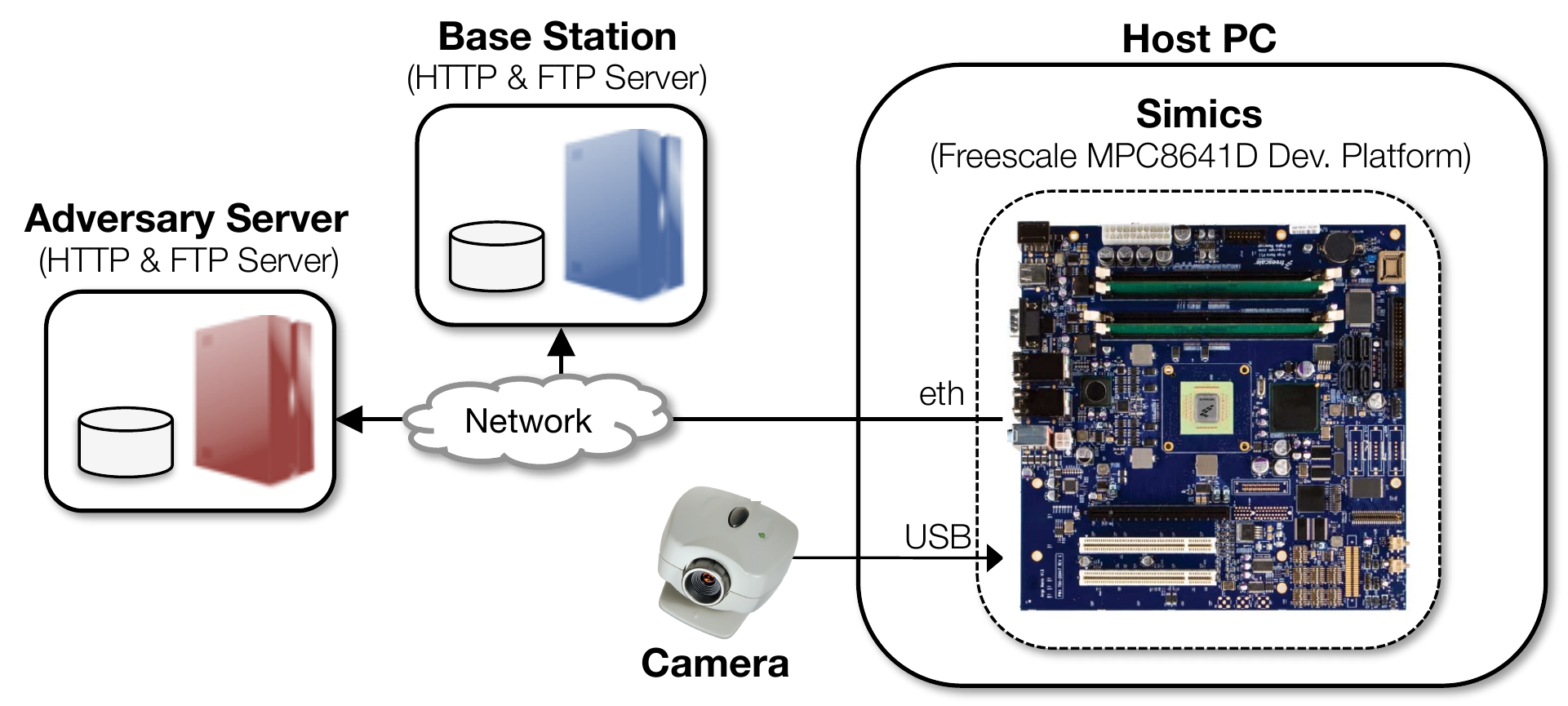}
\caption{The system setup for evaluation. The modified processor architecture (Figure \ref{fig:securecore} in Section \ref{sec::architecture}) is implemented on Simics.}
\label{fig:implementation}
\end{figure}

We implemented a prototype of our SCFD-based intrusion detection system on Simics~\cite{SimicsPaper}. Simics is a full-system simulator that can emulate a hardware platform including real firmware and device drivers, and also allows for processor micro-architecture modifications. Figure \ref{fig:implementation} shows the system setup used for our evaluation. We used the Freescale MPC8641D~\cite{mpc8641d} development platform on Simics. It has a dual-core processor, each core of which runs at 1350MHz and the system has a memory of $1 GB$ and runs an embedded Linux $2.6.23$. The SCTM was implemented by extending the \texttt{sample-user-decoder} on Simics. This allows us to implement the necessary ISA modification as described in Section \ref{sec::sctm}. The SPM has a total size of $8 KB$. 

%The implementation in this paper is not for evaluating the security, reliability of the system itself
%but for evaluating the intrusion detection method presented in \autoref{sec::analysis} and for 
%showing the feasibility of the architecture modifications from \autoref{sec::architecture}. 
%Thus, though different from the original SecureCore implementation~\cite{yoonsecurecore2013}, 
%we did not use/implement hypervisor and Simplex~\cite{sha01} because these are orthogonal to 
%the detection methods being presented here. However, it is always desirable for the final implementation
%to include these techniques.

\subsection{Target Application Model}
\label{subsec::application_model}

\begin{figure}[t]
\centering
\includegraphics[width=0.8\columnwidth]{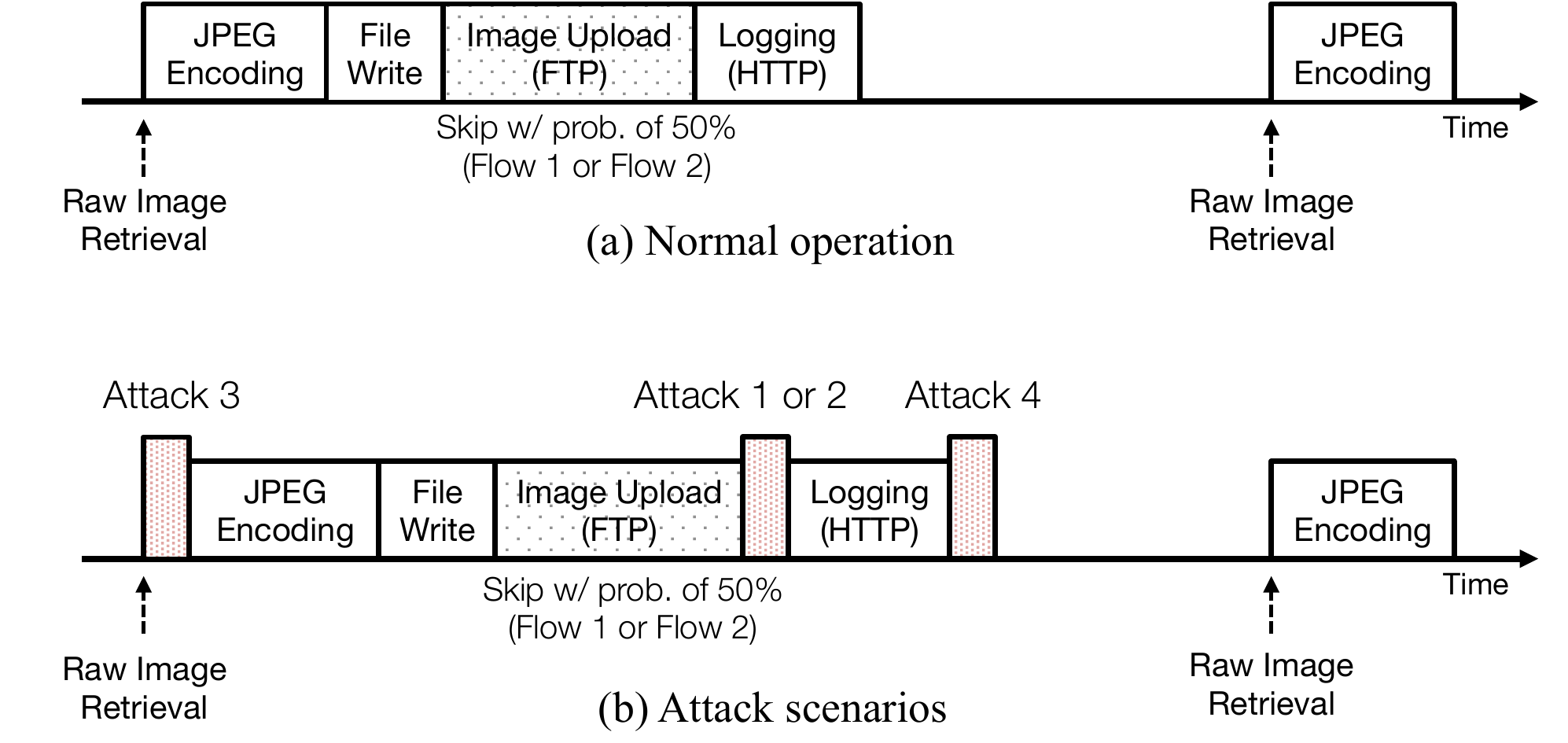}
\caption{The normal execution flow of the target application. \textsf{Attack 1} and \textsf{2} leak user authentication 
information and JPEG image through HTTP and FTP, resp., \textsf{Attack 3} erases the raw image buffer and \textsf{Attack 4} 
is a real-world shellcode that spawns a shell (\texttt{/bin/sh}) by executing \texttt{execve}~\protect\cite{shellcode89}.}
\label{fig:app_model}
\end{figure}

Figure \ref{fig:app_model}(a) shows the target application. Each invocation of the application (period of one 
second) cycles through the following steps: \ci retrieve a raw image from a camera, \cii compresses it to a JPEG 
format, \ciii upload the image file to the base station through FTP and finally {\em (iv)} write a log via  HTTP post. 
This type of application model (image capture $\rightarrow$ processing $\rightarrow$ communication) can be 
found in modern unmanned aerial vehicles (UAVs) that are used for surveillance or environmental 
studies~\cite{Kanistrasy:2013}. %Typically these systems have real-time constraints on their application tasks.

%Our system simulation platform, Simics, provides real-network connections that allow the application to access real HTTP and FTP servers (see \autoref{fig:implementation}). However, it does not provide USB connections but only serial port ones. Thus, we implemented a serial port daemon process running on the host that connects to the system simulator through a pseudo terminal. The daemon receives a command from the target application (that sees the connection as a real serial port) and captures a raw image from the USB camera and relays it to the application. 

\begin{figure}[t]
\centering
\includegraphics[width=0.6\columnwidth]{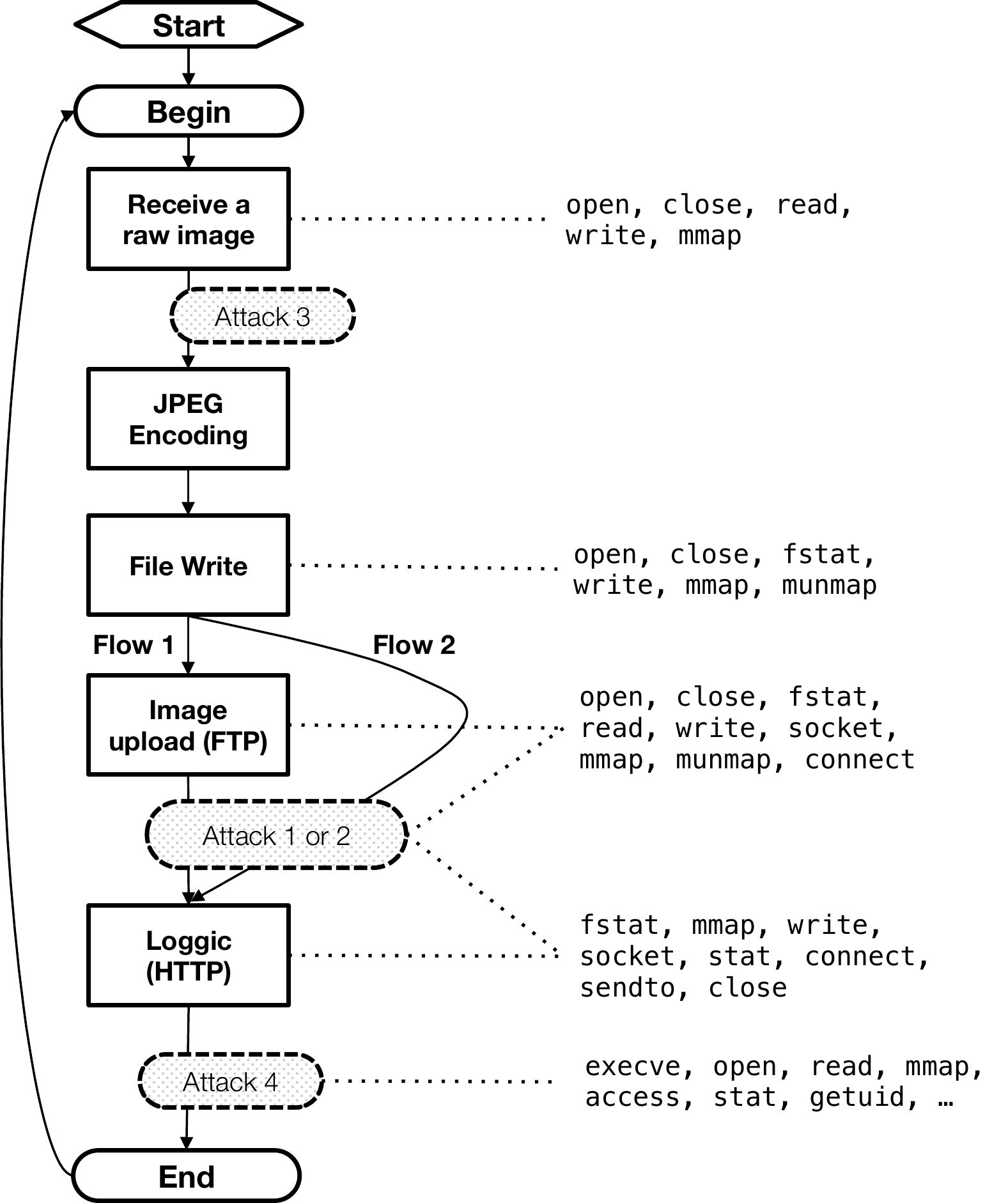}
\caption{The execution flow of the target application and the system call types used at each stage. Apart from those shown in the figure, the application used \texttt{futex}, \texttt{rt\_sigreturn} and \texttt{brk} system calls.} 
\label{fig:app_model_appendix}
\end{figure}

The distributions of the system call frequencies exhibited by this application is mainly affected by the stages after the JPEG compression. While the raw image size is always fixed (\eg $2.6 MB$ for $1280$ x $720$ resolution), a JPEG image size can vary ($27 KB$ -- $97 KB$) because of compression. This results in a 
variance in the number of \texttt{read} and \texttt{write} system calls. %To increase the complexity of the application and also to include further variations in the distribution of system calls, we added an additional code branch before the FTP upload stage that behaves as follows: the system could randomly skip the image upload process (based on a probability of $0.5$). 
To increase the complexity of the application and also to show certain scenarios that the proposed method cannot deal with well, we added an additional code branch before the FTP upload stage that behaves as follows: the system can randomly skip the image upload process (based on a probability of $0.5$). This affects the number of occurrences of network and file-related system calls during actual execution. Hence, the application has two legitimate flows, \textsf{Flow 1} and \textsf{Flow 2} as shown in Figure \ref{fig:app_model_appendix}; the figure also shows the system call types used at each stage of the execution flows.

We use this type of application model for the following reasons: \ca Simics, being a full system simulator that executes on a `host' system, is not fast enough to be able to control an actual system. Hence, we need to develop an application model that it can simulate; \cb we still need to demonstrate, in the simplest possible way, how our SCFD-based intrusion detection system works -- this application model is able to highlight the exact mechanisms and even its limitations. Note that this target application is crafted to show \emph{more} variance than many real embedded systems. Hence, if our detection method can catch changes in the system call distributions here then it can detect similar attacks in embedded systems that show less variance.

\subsection{Attack Scenarios}
\label{sec::attack_model}

We consider the following attack scenarios for this application:
\begin{enumerate}
\item \textsf{Attack~1} steals user authentication information used to connect to the base station's FTP server and sends it to an adversary HTTP server. {\em This attack invokes the same HTTP logging calls used by the legitimate executions}.

\item {\textsf{Attack~2}} uploads the image that was just encoded by the application to an adversary 
FTP server. {\em This attack also uses the same functions used by a legitimate FTP upload}.

\item {\textsf{Attack~3}} modifies the raw image array received from the camera. The attack erases the array by calling \texttt{memset}. {\em This attack does not require any system calls}. 

\item {\textsf{Attack~4}} is a real shellcode targeted at Linux on PowerPC and executes \texttt{execve} to spawn a shell (\texttt{/bin/sh})~\cite{shellcode89}. In general, a shellcode can be injected by data sent over a network or from a file and can be executed by exploiting buffer overflow or format string vulnerabilities. In our  implementation, the shellcode is stored in \texttt{char shellcode[]} and is executed by \texttt{\_\_asm\_\_("b shellcode")} when enabled. 
\end{enumerate}
The attack codes execute at spots marked in Figure~\ref{fig:app_model} when enabled. Note that our method is independent of where they happen since SCFDs do not care about the sequences of system calls.

\section{Evaluation Results}
\label{sec::eval_results}

We now evaluate the SCFD method on the prototype described in the previous section. We also compare it with an existing sequence-based approach to show how the two methods can be used to complement each other.

%\subsection{Results}
\subsection{Training}
\label{sec::eval_train}

To obtain the training set, we executed the system under normal conditions (\ie no attack present) 
 $2,000$ times. The target application used $14$ types of system calls (as shown in Figure \ref{fig:app_model_appendix}, together with \texttt{futex}, \texttt{rt\_sigreturn} and \texttt{brk}). We used the learning algorithm presented in Section \ref{sec::analysis} with settings, $\mathtt{MAX_K}=10$ and the total distance bound $\mathtt{Bound_{TD}}$ of $1,000$ on the resulting traces. %Since the application has a few control flow paths, it was safe to assume at most $10$ clusters. 
 The cutoff distance, $\theta$, which attests to the legitimacy of a new SCFD, is $1.95996$; \ie the significance level is $5\%$. We also
tested for $p_0=1\%$, \ie $\theta = 2.57583$. Again, the lower significance level, the more we are confident about the statistical significance when an outlier is observed. 

\vspace{\baselineskip}
Table \ref{tbl:syscalls} summarizes the training results. The first row shows the mean and the standard deviation 
of the SCFDs in the whole training set. The algorithm first reduce the dimensionality of the results from $14$ to $10$ by removing the system call types that show zero variance in the training set. The variations of \texttt{write} and \texttt{read} are due to the 
JPEG compression and the FTP uploading phases. The latter also affects the frequencies of the network- and file-related system call types. The global
$k$-means algorithm stopped at $k=5$ (the moment when the total distance becomes less than the bound $\mathtt{Bound_{TD}}$)
resulting in {\em five clusters} as shown in the same table. 

From these results, note first that the variation of each system call type is significantly reduced after the clustering; most of them become zero. This is because each cluster contains similar SCFDs, representing similar execution contexts. Also, from observing the mean values of the
system call types other than \texttt{write} and \texttt{read}, we can infer that Clusters $c_2$ and $c_3$ are from a similar execution 
context while the others are from a different context. We also observe that the former group corresponds to \textsf{Flow 1} because
of the additional system calls required for the FTP transfer. Also, the fewer number of \texttt{write} and \texttt{read} system calls of the second group suggest that they belong to \textsf{Flow 2}. As expected, within each group
clusters are distinguished by \texttt{write} and \texttt{read} due to the varying sizes of images that are compressed. The clustering results would be similar if $\mathtt{MAX_K}$ was set to, for example, $2$. In this case, one cluster would have the points from Clusters $c_1$, $c_4$ and $c_5$ combined but with different centroid and similarly $c_2$ and $c_3$ would constitute another new cluster. This could, however, blur boundaries between the execution contexts.

\begin{table}
\caption{The mean and the standard deviation of the system call frequency distributions in the entire training set 
and in each cluster after running the learning method. The boldfaced values represent the system call types whose variance is non-zero.\label{tbl:syscalls}}{\centering\scriptsize
\hspace*{-1.5cm}\begin{tabular}{|c|c|c||r|r|r|r|r|r|r|r|r|r||c|}
\hline & \# pts & & \texttt{write} & \texttt{read} & \texttt{mmap} & \texttt{open} & \texttt{close}& \texttt{fstat} & \texttt{munmap} & \texttt{socket} & \texttt{connect} & \texttt{stat} & Execution context\\
\hline
  \hline \multirow{2}{*}{All} & \multirow{2}{*}{2000} & Mean & \textbf{29.519}	& \textbf{101.197}	 & \textbf{1.520} & \textbf{2.514} & \textbf{4.548} & \textbf{1.520} & \textbf{1.520} & \textbf{2.034} & \textbf{2.034} & \textbf{4.034} &  \multirow{2}{*}{}\\
  \hhline{~~-----------}
    & & Stdev & \textbf{10.602} & \textbf{10.135} & \textbf{0.500} & \textbf{0.500} & \textbf{1.496} &  \textbf{0.500}  & \textbf{0.500} & \textbf{0.997} & \textbf{0.997} & \textbf{0.998}  &\\
  \hline 
  \hline 
  \multirow{2}{*}{$c_1$} & \multirow{2}{*}{490}& Mean & \textbf{17.376} &	91.000	&1.000&	2.000	&3.000	&1.000	&1.000	&1.000	&1.000&	3.000 & \multirow{2}{*}{\parbox{2.7cm}{Small image \\ No FTP upload}}\\
  \hhline{~~-----------}
  &  & Stdev & \textbf{1.246}	& 0.000 & 0.000	& 0.000	& 0.000	& 0.000	& 0.000	& 0.000	& 0.000	& 0.000  &\\
    \hline 
  \hline 
  \multirow{2}{*}{$c_2$} & \multirow{2}{*}{519}& Mean & \textbf{33.613} &	\textbf{108.306}&2.000&	3.000	&6.000	&2.000&	2.000	&3.000&	3.000	&5.000 & \multirow{2}{*}{\parbox{2.7cm}{Small-medium image \\ FTP upload}}\\
  \hhline{~~-----------}
   & & Stdev & \textbf{2.539} &	\textbf{1.269} & 0.000	& 0.000	& 0.000	& 0.000	& 0.000	& 0.000	& 0.000	& 0.000 & \\
     \hline 
  \hline 
  \multirow{2}{*}{$c_3$} & \multirow{2}{*}{506}& Mean & \textbf{43.708} &	\textbf{113.354}	&2.000&	3.000	&6.000	&2.000&	2.000	&3.000&	3.000	&5.000 & \multirow{2}{*}{\parbox{2.7cm}{Medium-large image \\ FTP upload}}\\
  \hhline{~~-----------}
  &  & Stdev & \textbf{4.539}&	\textbf{2.269} & 0.000	& 0.000	& 0.000	& 0.000	& 0.000	& 0.000	& 0.000	& 0.000  &\\
    \hline 
  \hline 
  \multirow{2}{*}{$c_4$} & \multirow{2}{*}{335}& Mean & \textbf{21.176} &	91.000	&1.000&	2.000	&3.000	&1.000&	1.000	&1.000&	1.000	&\textbf{2.998} & \multirow{2}{*}{\parbox{2.7cm}{Medium image \\ No FTP upload}}\\
  \hhline{~~-----------}
    & & Stdev & \textbf{1.080}	& 0.000 & 0.000	& 0.000	& 0.000	& 0.000	& 0.000	& 0.000	& 0.000	& \textbf{0.055} & \\
    \hline 
  \hline 
  \multirow{2}{*}{$c_5$} & \multirow{2}{*}{150}& Mean & \textbf{25.575} &	91.000	&1.000	&2.000&	3.000	&1.000	&1.000&	1.000&	1.000	&3.000 & \multirow{2}{*}{\parbox{2.7cm}{Large image \\ No FTP upload}}\\
  \hhline{~~-----------}
   & & Stdev & \textbf{1.627}	& 0.000 & 0.000	& 0.000	& 0.000	& 0.000	& 0.000	& 0.000	& 0.000	& 0.000  &\\
   \hline 
\end{tabular}}
\end{table}

\subsection{Accuracy}\label{sec::eval_eval}
Now, we evaluate the {\em accuracy} of our intrusion detection methods. We enabled each of the attacks from Section \ref{sec::attack_model}. 
For each attack type, we carried out $300$ execution instances and measured how many times the monitor detects malicious execution. 
An execution is considered malicious if any of the following is true: \ci any system call other than the $14$ observed types is detected; 
\cii any system call whose variance was zero during the profile ($4$ out of $14$ in the case above) actually exhibits variance or 
\ciii the distance of an observation from its closest cluster is longer than the threshold. Among these, rule \ci was never observed 
in the cases of {\textsf{Attacks 1-3}} because {\textsf{Attacks 1 \& 2}} re-used the same functions from normal executions and 
{\textsf{Attack 3}} makes no system calls at all. Table \ref{tbl:detection_results} summaries the results of our detection method 
as well as those of the sequence-based approach explained in Section \ref{subsec::eval_comparison}. The results of our SCFD method are as follows.

\begin{enumerate}%[leftmargin=0.4cm]

		\item {\textsf{Attack 1}}  (HTTP post): {\em All} the
		executions were classified as malicious based on rule \cii above since one 
		additional \texttt{brk} and \texttt{sendto}, each, were invoked. We tested the executions again, after removing such
		obvious situations. The results, however, did not change -- {\em all the malicious executions were caught by our monitor}. Because of the additional HTTP request by the attack code, \texttt{socket}, \texttt{connect}, 
		\texttt{close} and \texttt{stat} were called more in both \textsf{Flows 1} and \textsf{2}. System call types other 
		than the ones mentioned here were consistent with the profile. With $p_0=1\%$, the results were the same since 
		the use of the additional system calls already increased the distances from clusters to lie
		outside the acceptable boundaries.

		\item {\textsf{Attack 2}}  (FTP upload): If the attack code executes on \textsf{Flow 1}, it is easily caught because of the additional FTP transfer. As with the above, this attack changes some network related system calls. This is enough to make the SCFDs fall outside the 
		legitimate regions. The attack reads the image file as well. This increases the usage of \texttt{read} system calls thus further highlighting the anomalous behavior.

		On the other hand, if the attack is launched on \textsf{Flow 2} (that skips the FTP upload to the base station), it may not be as easy to detect. Since the attack uses the same functions that are invoked by legitimate code, it looks 
		like the application is following \textsf{Flow 1} (where the FTP upload is actually legitimate). In this case, only 
		$1\%$ of the malicious code executions were caught. The detection rates would be significantly higher if the attacker either used 
		different images that vary in size or used code that utilizes different combinations of system calls. The latter case would
		also hold for the HTTP post attack. {\bf Note}: the detection was not successful because we \emph{tailored} the attack instance to {\em closely match} legitimate execution (especially due to our knowledge of the detection methods); however, many attacks will not be able to match legitimate execution in such a precise manner and will end up being caught. %while it looks like the detection mechanism was not successful in this case, that is only because we \emph{tailored} the attack instance to {\em closely match} legitimate execution (especially due to our knowledge of the detection methods); many attacks will not be able to match legitimate execution in such a precise manner and will end up being caught.

		\item {\textsf{Attack 3}} (Data corruption): This attack does not use any system calls; it just changes the values of
		the data. However, this may affect the execution of code segments that follow, especially ones that depend on the data -- the JPEG compression. The attack code resets the raw image data by using \texttt{memset}. This is 
		compressed by the JPEG encoder that produces $15$ KB of black images. This attack was {\em always} caught by our monitor because these image sizes are not typical during normal execution. Hence, calls to \texttt{read} and 
		\texttt{write} were much less frequent when compared to normal execution. % (where these calls were used often to write the larger images to files or upload on FTP servers). 
		The attack could have circumvented our detection method if, \eg 
		the raw image is just replaced with another that has a similar after-compression size as the original or only a part 
		of the image is modified.\footnote{If the attacker modified the compressed image, our method cannot detect it because
		the system call usage would never change. This, however, does not fall into our threat model explained in 
	        Section \ref{sec::intro}.} But performing either of these actions may also trigger the use of additional system calls
		that would be caught by our monitor.
		
		\item {\textsf{Attack 4}} (Shellcode execution): This attack was easily detected since it uses \texttt{execve} (used by the shellcode) which was never observed during the profiling phase. Furthermore, it was followed by a bunch of system calls including \texttt{open}, 
		\texttt{mmap}, \texttt{access}, \texttt{getuid}, \textit{etc}. This was due to the execution of a shell, 
		\texttt{/bin/sh}, spawned by the injected \texttt{execve}. In fact, \texttt{INST\_END} was not executed since \texttt{execve} 
		does not return on success. Nevertheless, the attack could be detected because a watchdog timer was used to \emph{wake up}
		the secure monitor that then checks the application's SCFD traced until the timer expires. From this experiment it can be 
		expected that our method can detect more sophisticated shellcode \cite{shellcode801} that uses unusual system calls, \eg \texttt{setreuid}, \texttt{setregid}, \textit{etc.}\footnote{The shellcode used in our experiment is simpler than 
		the ones targeted for Linux/x86~\cite{shellcode801} due to the scarcity of sophisticated shellcode for Linux/PowerPC. } 
\end{enumerate}

\begin{table}
\caption{Comparison between \textsf{SCFD} and \textsf{PST} (a sequence-based) methods.\label{tbl:detection_results}}{\centering
\begin{tabular}{|c||c|c|c|}
  \hline Type  & \textsf{SCFD} & \textsf{PST} & Cause \\
  \hline
  \hline \parbox{1.5cm}{Attack 1} & $100\%$ & $100\%$ & \parbox[c][3em][c]{9cm}{\textsf{SCFD}: Extra network-related system calls\\\textsf{PST}: Unusual transition (HTTP-HTTP)}\\
  \hline \parbox{1.5cm}{Attack 2\\ (Flow 1)} & $100\%$ & \parbox{2.5cm}{$0\%$ ($\bar{N}=3$)\\ \\$100\%$ ($\bar{N}=5$)} & \parbox[c][4em][c]{9cm}{\textsf{SCFD}: Extra network- and file-related system calls\\\textsf{PST} : Short sequence cannot capture unusual transition (FTP-FTP)}\\

  \hline \parbox{1.5cm}{Attack 2\\ (Flow 2)} & $1\%$ & \parbox{2.5cm}{$0\%$ (both $\bar{N}$)} & \parbox[c][2em][c]{9cm}{Both: Not differentible from legitimate Flow 1}\\
  \hline \parbox{1.5cm}{Attack 3} & $100\%$ & \parbox[c]{2.5cm}{$0\%$ ($\bar{N}=3$)\\ \\ $100\%$ ($\bar{N}=5$)} & \parbox[c][3em][c]{9cm}{\textsf{SCFD}: Too small image size\\\textsf{PST}: Short sequence cannot capture shortened \texttt{write} chain}\\
  \hline \parbox{1.5cm}{Attack 4} & $100\%$ & $100\%$ & \parbox{9cm}{Both: \texttt{execve} was never seen}\\
  \hline
\end{tabular}}
\end{table}

\noindent \textbf{False Positives}: The {\em false positive rate} is just as important as the detection rate because frequent false alarms degrade system availability. 
To measure the false positive rates, we obtained a new set of SCFDs by running the system without activating any attacks and measured how many times 
the secure monitor classifies an execution as being suspicious. Most false positives in these tests were due to the images sizes that, when 
compressed, fell below the normal ranges. For the cut-off distance $\theta$ with $p_0=5\%$, $35$ out of $2,000$ executions ($1.75\%$) 
were classified as malicious. With $p_0=1\%$, \ie a farther cut-off distance, it was reduced to just $17$ ($0.85\%$). Such a lower significant
level relaxes the cutoff distance and produces fewer false alarms because even some rarely-seen data points are considered normal. However,
this may result in lower detection rates as well. In the attack scenarios listed above, however, the results did not change even with $p_0=1\%$. 
This is a consideration for system designers to take into account when implementing our intrusion detection methods; they will have a better feel for
when certain executions are normal and when some are not. Hence, they can decide to adjust values for $p_0$ based on the actual system(s) 
being monitored.\\

These results show that our method can effectively detect malicious execution contexts without relying on complex analysis. While it is true
that the accuracy of the method may depend on the attacks that are launched against the system, in reality an attacker would need to not only know 
the exact distributions of system call frequencies but also be able to implement an attack with such a limited set of calls -- both of these
requirements significantly raise the difficulty levels for would-be attackers.

\subsection{Comparison with Sequence-based Approach}
\label{subsec::eval_comparison}

To show how our detection method can effectively complement existing system call-based intrusion detection methods, we compare it against a sequence-based approach using the same data set used in the previous section. Among the existing methods (explained in Section \ref{sec::related}), we use a variable-order Markovian model (VMM) \cite{Begleiter:2004} using \emph{Probabilistic Suffix Tree} (\textsf{PST})~\cite{Sun:2006,Chandola:2012}, due to its ability to learn significant sequence patterns of different lengths. This enables us to calculate $\Pr(s_{(t)} |$ $s_{(t-N)}$ $\cdots$ $s_{(t-2)}$ $s_{(t-1)})$, that is, the probability of a system call made at time $t$ given a recent history, without having to learn all or a fixed-length of the sequences (used by $N$-gram or fixed-order Markovian models). The sequence length, $N$, varies with different patterns and is learned from their significance in the training set. 
%Among the existing methods (explained in Section \ref{sec::related}), we use a variable-order Markovian model using \emph{Probabilistic Suffix Tree} (\textsf{PST})~\cite{Chandola:2012,Sun:2006}, due to its ability to learn a set of subsequences of different lengths, each of which is a significant pattern. This enables us to calculate $\Pr(s_{(t)} |$ $s_{(t-N)}$ $\cdots$ $s_{(t-2)}$ $s_{(t-1)})$, that is, the probability of a system call made at time $t$ given a recent history, without having to learn all or a fixed-length of the history (used by $N$-gram or fixed-order Markovian models). The history size, $N$, varies with different patterns and is learned from their significance in the training set. 

\textsf{PST} learns the conditional probabilities in a (suffix) tree structure and thus requires a user defined parameter $\bar{N}$ that limits the maximum depth of the tree, i.e., the length of sequence pattern. We tested two configurations: \ci $\bar{N}=3$ and \cii $\bar{N}=5$ that show different results. Given the \textsf{PST} learned from the training set, a test is carried out as follows: for each system call, we calculate its (conditional) probability using the \textsf{PST} and consider it malicious if the probability is less than a threshold of~$1\%$. An iteration (\ie one execution) of the target application is classified as being malicious if any call is classified as malicious. Table \ref{tbl:detection_results} summaries the results from the detection techniques.

\begin{enumerate}

\item {\textsf{Attack~1}} (HTTP post): \textsf{PST} was able to detect \textsf{Attack~1}
because of this particular sequence: \texttt{sendto}-\texttt{close}-\texttt{write}-\texttt{write},
where we have that $\Pr(\mathtt{wr|se-cl-wr}) = 0$
since in the legitimate executions, \texttt{sendto}-\texttt{close}-\texttt{write}, the end of the HTTP logging function, is almost always followed by \texttt{stat}. The last \texttt{write}, which is the beginning of the HTTP function caused an \emph{unusual transition} and hence detected by \textsf{PST}. On the other hand, \textsf{SCFD} detected the attack because of the \emph{unusual frequencies} of network-related system calls.

\item {\textsf{Attack~2}} (FTP upload):\footnote{Some of the executions included FTP server error (due to too many connections) that caused the malicious FTP session to be disconnected. Both \textsf{SCFD} and \texttt{PST} classified such executions to be malicious and hence we excluded them from the accuracy results.} In normal executions of Flow 1, the legitimate FTP operation ends with this sequence: \texttt{write}-\texttt{close}-\texttt{close}-\texttt{munmap}-\texttt{write}. Then, the legitimate HTTP operation starts with a \texttt{write} call. The extra FTP operation by \textsf{Attack 2}, which executes between the two operations, starts with \texttt{socket}. With $\bar{N}=5$, \textsf{PST} was able to detect the malicious executions because of the unusual sequence. However, with $\bar{N}=3$, it was \emph{not} able to detect the extra FTP operation on Flow 1 at all. This is because the sequence \texttt{close}-\texttt{munmap}-\texttt{write}-\texttt{socket} is a legitimate one used at the boundary between File Write and the FTP upload stages (which is described in Figure \ref{fig:example_new} in Section \ref{sec::intro}). That is, the short sequence of calls was not able to differentiate the sequence from the ones generated by the transition between the legitimate FTP and the extra FTP stages. For Flow 2, \textsf{PST} was not able to detect the attacks at all for any $\bar{N}$ (in addition to $3$ and $5$). This is because, as explained in Section \ref{sec::eval_eval}, these executions are identical to the legitimate executions on Flow 1. Hence, both \textsf{SCFD} and \textsf{PST} were not able to catch the attacks.

\item \textsf{Attack~3} (Data corruption): This attack only alters the number of \texttt{read} and \texttt{write} 
calls. In particular, the size of modified images affects the length of \texttt{write} chain in \texttt{mmap}-\texttt{write}-$\cdots$-\texttt{write}-\texttt{close} sequence made
when writing the compressed JPEG image to a file (see `File Write' stage in Figure \ref{fig:example_new}). When an attacker modifies the image, the length of the \texttt{write} chain becomes shorter. However, \textsf{PST} with a short sequence ($\bar{N}=3$) was not able to classify any of the executions as being malicious because the length of the chain becomes 4 when the image is corrupted. Hence, the malicious executions were detected with $\bar{N}=5$, because the probability of seeing \texttt{close} after \texttt{mmap}-\texttt{write}-\texttt{write}-\texttt{write}-\texttt{write} is zero given the training set. Note that, if the image size got larger instead and thus made the \texttt{write} chain longer than usual, sequence-based methods cannot detect this behavior because the only change would be that there are more chains of \texttt{write} that have a legitimate length. \textsf{SCFD}, on other hand, can detect this type of attack easily.

\item \textsf{Attack 4} (Shellcode execution): \textsf{Attack 4} was easily caught by both methods because \texttt{execve} call was never seen in the normal trace. 

\end{enumerate}

\vspace{\baselineskip}
The results suggest that sequence-based approaches (the \textsf{PST} method in our evaluation) are sensitive to local, temporal variations, \eg an unusual transition from \texttt{write} to \texttt{write} instead of to \texttt{stat}. Our \texttt{SCFD} might not catch such a small, local variation.\footnote{In fact it can because having the extra \texttt{stat} is a strong indicator of anomaly by our normal execution model shown in Table \ref{tbl:syscalls}.} However, sequence-based approaches fail to detect abnormal deviations in high-level, naturally variable execution contexts such as network activities or diverse data. This is because these require a global view on the frequencies of different system call types made during an entire execution. Hence, one can use these two approaches together to improve the overall accuracy of the system call-based IDS for embedded systems. From the implementation perspective, one may apply a sequence-based method to each system call observed and at the same time create an SCFD during an entire execution. Then, at the execution boundary, the SCFD method can be applied to check if the high-level execution context is anomalous or not. 

\subsection{Time Complexity}
\label{subsec::overhead}
%\noindent \textbf{Time Complexity}: 

To evaluate the time complexity of the proposed detection method, we measured the number of instructions retired by the function that finds 
the closest cluster (Eq.~\eqref{eq:closest_cluster}) among the five clusters given a new observation and the average time to perform the 
analysis.\footnote{Simics is not a cycle-accurate simulator. Thus, the times are measured on a real machine with Intel Core I5 1.3GhZ dual-core
processor. The analysis code is compiled with $-O0$ option. The statistic is based on $10,000$ samples.}
 
\begin{table}[t]
\caption{Time Complexity of our Analysis\label{tab::complexity}}\centering{
\begin{tabular}{|c||c|c|}
  \hline \# of system  & Number of& Avg. (Stdev.) of\\
  call types & instructions &  analysis times\\
  \hline
  \hline 5 & 2175 & 0.914 $\mu s$ (0.553 $\mu s$)\\
  \hline 10 & 4875 & 2.624 $\mu s$ (1.405 $\mu s$)\\
  \hline 14 & 8125 & 5.231 $\mu s$ (1.965 $\mu s$)\\
  \hline
\end{tabular}}
\end{table}
As Table \ref{tab::complexity} shows, the detection process is fast. This is possible because we store $\Sigma^{-1}$, the inverse of 
the covariance matrix, of each cluster, not $\Sigma$. A Mahalanobis distance is calculated in $\mathcal{O}(D^2)$, where $D$ is the number of
system call types being monitored, since in $(\mathbf{x}^*-\boldsymbol\mu)^T \boldsymbol\Sigma^{-1} (\mathbf{x}^*-\boldsymbol\mu)$, 
the first multiplication takes $\mathcal{O}(D^2)$ and the second one takes $\mathcal{O}(D)$. Note that it would have taken $\mathcal{O}(D^3)$ if we stored the covariance 
matrix itself instead of its inverse; since a $D\times D$ matrix inversion takes $\mathcal{O}(D^3)$. 

Note, again, that the monitoring and
detection methods are \emph{not in the critical path}, \ie they do not affect the execution of the applications we monitor since they are 
\emph{offloaded} onto the secure core. More importantly, the time complexity of our method is \emph{independent} of how often and many times
the application uses system calls; it only depends on the number of system call types being monitored. This is determined in the training phase 
and does not change during the monitoring phase (see Section \ref{subsec::dimension}). On the other hand, the overheads of sequence-based approaches are highly dependent on the application complexity (\ie how many system calls are made). Hence, the deterministic time complexity of our SCFD method makes it particularly suitable for embedded systems. 

\subsection{Limitations and Possible Improvements}
\label{subsec::limitations}

One of the limitations of our detection algorithm is that it checks for intrusions \emph{after} execution is complete (at least for that
invocation). Thus, if an attack tries to suddenly break the system, we cannot detect or prevent it. Combining a sequence-based method with our \textsf{SCFD} can be a solution if such attacks can be detectable by the former. If not, one can increase the chances of 
detection such problems by splitting the whole execution range into blocks \cite{yoonsecurecore2013} and checking for the distribution of 
system calls made in each block as soon as the execution passes each block boundary. This, however, would need more computation in the
secure core at run-time, more storage in the SPM and a few more code modifications.

Another way to handle this problem is to combine this analysis/detection with other behavioral signals, especially ones that have a
finer granularity of checks, \eg timing~\cite{yoonsecurecore2013}. Since some blocks may use very few
system calls (perhaps none) or even a very stable subset of such calls we can monitor the execution time spent in 
such a block to reduce the SCFD-based overheads (which is still low). This keeps the profile from bloating and 
prevents the system from having to carry out the legitimacy tests. We can also use the timing information in 
conjunction with the system call distribution; \ie by learning the normal time to execute a distribution of system
calls, we can enforce a policy where each application block executes all of its system calls within (fairly)
tight ranges. This is, of course, provided that the system calls themselves do not show unpredictable timing
behavior. This makes it much harder for an attacker who imitates system calls \cite{Wagner:2002} or one that replaces certain system calls with malicious functions~\cite{syscall_hijacking}.

\vspace{\baselineskip}
Our model represents the histograms of system calls. In fact, each signature (histogram) would be represented by a mixture of multinomial distributions. That is, our model assumes generative models which select a mixture of multinomial parameters and then generate a histogram of system calls. Here, we assume that each multinomial distribution can be accurately represented by the multivariate Gaussian distributions (the multivariate Gaussian approximation of multinomial for large numbers of samples).
In this regard, the assumption on our model is closely related to the \emph{topic model} such as Latent Dirichlet Allocation \cite{Blei:2003:LDA,Blei:2012:PTM}. Here, we build a pragmatic lightweight module.
One of the main drawbacks of the $k$-means clustering algorithm is that one may need to know or predefine the number of clusters. That is, system behaviors should be correctly represented by $k$ numbers of multinomial (Gaussian) distributions of histogram. Some large-scale systems would have many heterogeneous modes (distributions). In this case, the appropriate solutions would be using non-parametric topic models such as Dirichlet process. However, we empirically observed that many embedded systems with predictable behavior can be represented by a tractable number of clusters. Thus, we use a simpler model with the $k$-means cluster.

%The current approach for 

%Some systems may exhibit highly unpredictable, but yet legitimate, memory usage caused by, for example, network activities or user interactions. In these cases, our current model may alarm many false positives. To deal with such problems, we plan to build a robust classification algorithm by extracting local features from MHMs in an unsupervised manner as in Deep Learning 

%TODO: talk about .... our technique is limited to applications that are quite restricted.... we will... execution context mining using... deep learning...

%% file: related.tex
\section{Related Work}
\label{sec::related}

Forrest \textit{et al.}~\cite{Forrest:1996} build a database of look-ahead pairs of system calls; for each system call type, what is the next $i^{th}$ system call for $i=1,2,$ up to $N$. Then, given a longer sequence of length $L>N$, the percentage of mismatches is used as the metric to determine abnormality. Hofmeyr \textit{et al.}~\cite{Hofmeyr:1998} extends the method by profiling unique sequences of \emph{fixed} length $N$, called an \emph{N-gram}, to reduce the database size. The legitimacy test for a given sequence of length $N$ is carried out by calculating the smallest Hamming distance between it and the all sequences in the database. %This fundamental model evolved in various aspects in follow-up work. For example, 
The N-gram model requires a prior assumption on suitable $N$ because it affects the accuracy as well as the database size. Marceau \cite{Marceau:2000} proposes a finite state machine (FSM) based prediction model to relax these requirements and Eskin \textit{et al.}~\cite{Eskin01modelingsystem} further improves by employing a \emph{wild-card} for compact sequence representation. Markovian techniques such as Hidden Markov model (HMM) \cite{Warrender99detectingintrusion} and %, Markov chains \cite{Jha:2001} 
variable-order Markov chain \cite{Sun:2006} have also been explored. Chandola \textit{et al.}~\cite{Chandola:2012} provides an extensive survey on various anomaly detection techniques for discrete sequences. A similar approach to our work is \cite{Burguera:2011}, in which the system call counts of Android applications (traced by a software tool called \emph{strace}) are used to find malicious apps. Using a crowdsourcing, the approach collects the system call counts of a particular application from multiple users and applies $k$-means (with Euclidean distance metric) to divide them into \emph{two} clusters. A smaller cluster is considered to be malicious based on the assumption that benign apps are the majority.

There has also been work on system call arguments monitoring. Mutz \textit{et al.}~\cite{Mutz:2006} introduce several techniques to test anomalies in argument lengths, character distribution, argument grammar, \textit{etc.} Maggi \textit{et al.} \cite{Maggi:2010} use a clustering algorithm to group system call invocations that have similar arguments. 

As previously mentioned, the usual way of system call instrumentation relies on an audit module in the OS layer. 
Hardware-based system call monitoring mechanism can improve the overall security of the 
system by cutting off a potential vulnerability -- the software audit module. Pfoh \etal \cite{Pfoh:2011} proposed Nitro, a hardware-based system call tracing system where system calls made inside virtual machines in manner similar to ours (Section~\ref{sec::sctm}). We note that our detection method (Section~\ref{sec::analysis}) is orthogonal to how system calls are traced. Hence we can implement it on systems like Nitro. Other types of instrumentation include static analysis of program source code \cite{Wagner:2001} and user-level processes for system call interposition \cite{Jain:1999}.

The SecureCore architecture~\cite{yoonsecurecore2013} takes advantage of the \emph{redundancy} of 
a multicore processor; a \emph{secure core} is used to monitor the run-time execution behavior of target applications running on a \emph{monitored core}.  The original architecture is designed to watch applications \emph{timing} behavior. \cite{YoonMHM:2015} extends the architecture by building a memory behavior monitoring framework for system-wide anomaly detection in real-time embedded systems. There also exists some work in which a multicore processor (or a coprocessor) is employed as a security measure, such as \cite{Chen:2008,Kannan:2009,Deng:2010,embeddedsecurity:s3a2012,Lo:2014} for instruction-grain monitoring.

%,

%------------------
%
%A Comparison with Gaussian Kernel Density Estimator
%
%-----------------
%
%% Jaesik: you can put this either in introduction or related work.
%
%Our method includes improvements over the recent non-parametric method \cite{Securecore2012} in two key aspects. The (Gaussian) kernel density estimator is modeled on one-dimension (time of executions of a computational unit). This method enables modeling system call distributions which are collected from multi-dimensions features.  Another main drawbacks of non-parametric methods are that the number of parameters increases as training data increase. Data reduction methods such as collapsed Gaussian mixture \cite{Goldberger2005} handle the issues by the bottom-up clustering from individual Gaussians to a small numbers of clusters. Here, the global k-means clustering algorithm also reduces the number of parameters by the top-down approach.
%
%-----------------
%
%

%% file: concl.tex
\section{Conclusion}
\label{sec::concl}

In this paper we presented a lightweight intrusion detection method that uses application execution contexts learned from system call frequency distributions of embedded applications. We demonstrated that the proposed detection mechanism could effectively complement sequence-based approaches by detecting anomalous behavior due to changes in high-level execution contexts. We also proposed certain architectural modifications to aid in the monitoring and analysis process. The approaches presented in the paper are limited in terms of the target applications and demonstration. Hence, as future work, we intend to implement the proposed architecture on a soft processor core~\cite{Leon3} and to evaluate our method with real-world applications. We also plan to improve the learning and analysis methods using the topic modeling approach (explained in Section \ref{subsec::limitations}) to deal with large-scale heterogeneous behaviors of complex embedded applications.

%we plan to combine the SCFD approach with the timing-based approach \cite{yoonsecurecore2013} as described in Section \ref{subsec::limitations}. 
%Also, we intend to implement the proposed architecture on a soft processor core~\cite{microblaze} and to evaluate our method with real-world applications.

%% file: appendix.tex
% Appendix

\section*{APPENDIX}

\setcounter{section}{1}
%In this appendix, 
\appendix

\section{Cutoff Distance} \label{appendix::cutoff}
%
%First, let $z$ be a Mahalanobis distance from a multivariate normal distribution. Then, 
%\begin{align}\label{eq:cutoff}
%\int_{0}^{\theta} c \cdot  e^{-\frac{1}{2}z^2}dz = 1 - p_0, 
%\end{align}
%where $c$ is a normalizing constant that satisfies Equation \eqref{eq:cutoff} with $\theta=\infty$ and $p_0=0$ by the 
%definition of a probability density function (PDF). This results in $c=1/1.25331$ because 
%\begin{align}
%\int_{0}^{\infty}  e^{-\frac{1}{2}z^2}dz \backsimeq \Big[1.25331\cdot \mathrm{erf}(0.707107\cdot z)\Big]^{\infty}_{0} = 1.25331,\nonumber
%\end{align}
%where $\mathrm{erf}(z)$ is the \emph{error function} and is $1$ and $0$ for $z=\infty$ and $z=0$, respectively. 
%Accordingly, \eqref{eq:cutoff} becomes
%\begin{align}
%\frac{1}{1.25331}\int_{0}^{\theta} e^{-\frac{1}{2}z^2}dz &\backsimeq \frac{1}{1.25331}\Big[1.25331 \cdot \mathrm{erf}(0.707107\cdot z)\Big]^{\theta}_{0}\nonumber\\
%&= \mathrm{erf}(0.707107\cdot \theta) = 1 - p_0.\nonumber
%\end{align}
%Therefore, the cutoff distance $\theta$ for a significant level $p_0$ is
%\begin{align}\label{eq:theta}
%\theta = \frac{\mathrm{erf}^{-1}(1-p_0)}{0.707107}.
%%\theta = {\mathrm{erf}^{-1}(1-p_0)}/{0.707107}.
%\end{align} 
%For $p_0=1\%$ and $5\%$, $\theta \sim 2.57583$ and $1.95996$, respectively. 
%

In general, there is no analytic solution for calculating the cumulative distribution function (CDF) for  multivariate normal distributions. However, it is possible to derive the CDF with Mahalanobis distance. The cutoff distance $\theta$ can be derived by finding the smallest distance that makes the probability that a data point $\mathbf{x}$, which in fact belongs to the cluster and has a distance farther than $\theta$, is not greater than $p_0 = 0.01$ or $0.05$.

\begin{figure}[b]
\centering
\includegraphics[width=0.4\columnwidth]{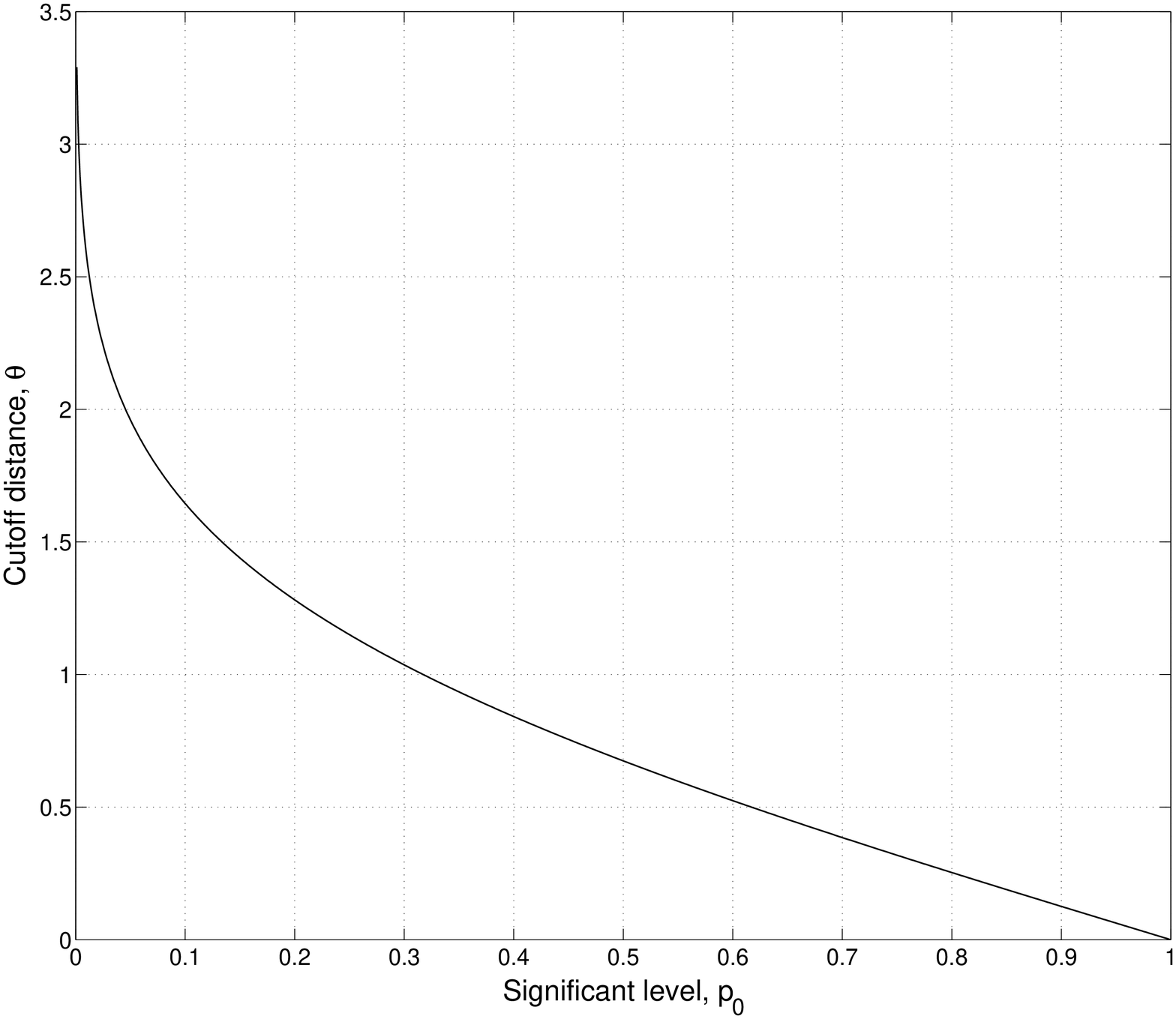}
\caption{The cutoff distance $\theta$ for significant level $p_0$.}
\label{fig:cutdist}
\end{figure}

First, let $z$ be a Mahalanobis distance from a multivariate normal distribution. Then, 
\begin{align}\label{eq:cutoff}
\int_{0}^{\theta} c \cdot  e^{-\frac{1}{2}z^2}dz = 1 - p_0, 
\end{align}
where $c$ is a normalizing constant that satisfies Eq. \eqref{eq:cutoff} with $\theta=\infty$ and $p_0=0$ by the 
definition of a probability density function. This results in $c=1/1.25331$ because 
\[\int_{0}^{\infty}  e^{-\frac{1}{2}z^2}dz \backsimeq \Big[1.25331\cdot \mathrm{erf}(0.707107\cdot z)\Big]^{\infty}_{0} = 1.25331,\]
where $\mathrm{erf}(z)$ is the \emph{error function} and is $1$ and $0$ for $z=\infty$ and $z=0$, respectively. 
Accordingly, Eq. \eqref{eq:cutoff} becomes  
\begin{align*}
\frac{1}{1.25331}\int_{0}^{\theta} e^{-\frac{1}{2}z^2}dz &\backsimeq \frac{1}{1.25331}\Big[1.25331 \cdot \mathrm{erf}(0.707107\cdot z)\Big]^{\theta}_{0}\nonumber\\
&= \mathrm{erf}(0.707107\cdot \theta) = 1 - p_0.
\end{align*}
Therefore, the cutoff distance $\theta$ for a significant level $p_0$ is
\begin{align}\label{eq:theta}
\theta = \frac{\mathrm{erf}^{-1}(1-p_0)}{0.707107}.
%\theta = {\mathrm{erf}^{-1}(1-p_0)}/{0.707107}.
\end{align} 
For $p_0=1\%$ and $5\%$, $\theta \sim 2.57583$ and $1.95996$, respectively. Figure~\ref{fig:cutdist} shows the cutoff distance for $0\%\le p_0\le 100\%$. The cutoff distance is not bounded (\ie $\theta=\infty$) when $p_0=0\%$ and is $0$ when $p_0=100\%$.